\documentclass[fleqn,usenatbib]{rasti}

\usepackage{newtxtext,newtxmath}

\usepackage[T1]{fontenc}
\usepackage{ulem}
\DeclareRobustCommand{\VAN}[3]{#2}
\let\VANthebibliography\thebibliography
\def\thebibliography{\DeclareRobustCommand{\VAN}[3]{##3}\VANthebibliography}

\usepackage{graphicx}	
\usepackage{amsmath}	
\usepackage{xcolor}

\title[Mitigating modal noise]{Mitigating Modal Noise in Multimode Circular Fibres by Optical Agitation using a Galvanometer}

\author[S. Ghosh et al.]{
Supriyo Ghosh,$^{1}$ \thanks{E-mail: s.ghosh3@herts.ac.uk, supriyoani89@gmail.com}
Chantira Boonsri,$^{2}$
William Martin,$^{1}$
Hugh R. A. Jones,$^{1}$
Piyamas Choochalerm,$^{2}$
\newauthor
Sarah Usher,$^{3}$ 
Stephanos Yerolatsitis,$^{4,5}$
Thomas Wocial,$^{1}$
and Thomas Wright $^{4}$
\\
$^{1}$University of Hertfordshire, Hatfield, Hertfordshire, AL10 9AB, UK\\
$^{2}$National Astronomical Research Institute of Thailand (NARIT),  Don Kaeo, Mae Rim District, Chiang Mai 50180, Thailand\\
$^{3}$ Thorlabs Vytran Europe, 2 Kew Court, Exeter, EX2 5AZ, UK\\
$^{4}$ University of Bath, Claverton Down, Bath, BA2 7AY, UK \\
$^{5}$ University of Central Florida, 4000 Central Florida Blvd. Orlando, Florida, 32816, USA\\
}

\date{Accepted XXX. Received YYY; in original form ZZZ}

\pubyear{2023}

\begin{document}
\label{firstpage}
\pagerange{\pageref{firstpage}--\pageref{lastpage}}
\maketitle

\begin{abstract}
Modal noise appears due to the non-uniform and unstable distribution of light intensity among the finite number of modes in multimode fibres. It is an important limiting factor in measuring radial velocity precisely by fibre-fed high-resolution spectrographs. The problem can become particularly severe as the fibre’s core become smaller and the number of modes that can propagate reduces. Thus, mitigating modal noise in relatively small core fibres still remains a challenge. We present here a novel technique to suppress modal noise. Two movable mirrors in the form of a galvanometer reimage the mode-pattern of an input fibre to an output fibre. The mixing of modes coupled to the output fibre can be controlled by the movement of mirrors applying two sinusoidal signals through a voltage generator. We test the technique for four multimode circular fibres: 10 and 50 $\mu$m step-index, 50 $\mu$m graded-index, and a combination of 50 $\mu$m graded-index and 5:1 tapered fibres (GI50t). We present the results of mode suppression both in terms of the direct image of the output fibre and spectrum of white light obtained with the high-resolution spectrograph. We found that the galvanometer mitigated modal noise in all the tested fibres, but was most useful for smaller core fibres. However, there is a trade-off between the modal noise reduction and light-loss. The GI50t provides the best result with about 60$\%$ mitigation of modal noise at a cost of about 5$\%$ output light-loss. Our solution is easy to use and can be implemented in fibre-fed spectrographs.

\end{abstract}




\section{Introduction}

\begin{figure*}
	\centering
	\includegraphics[width=6.5in, height=2.5in]{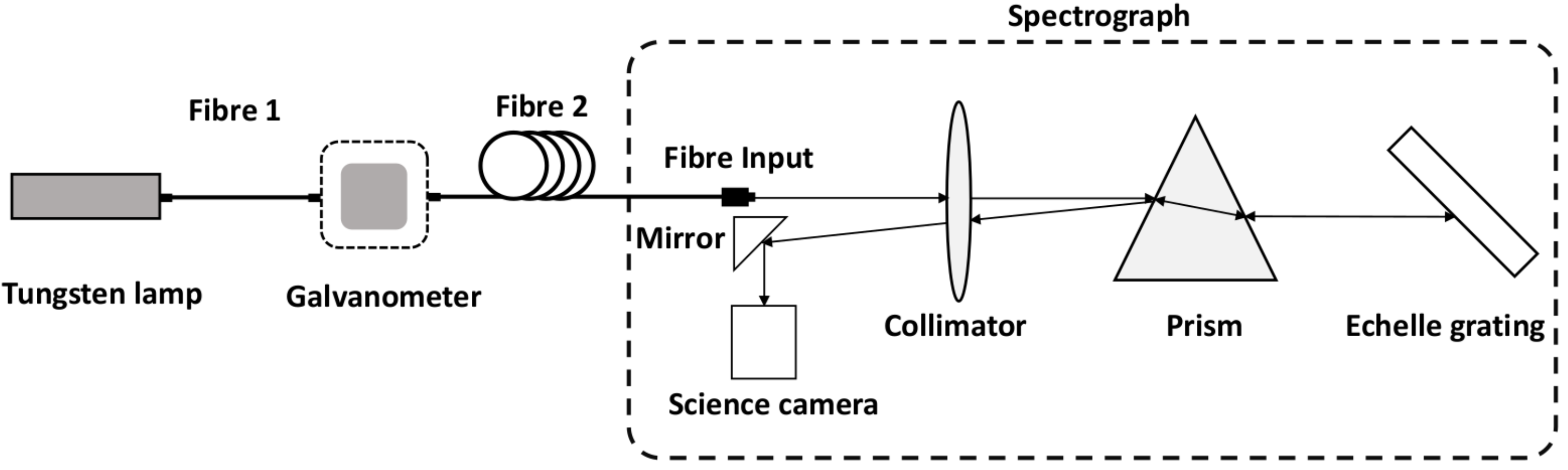}
 	\caption{Schematic layout of the experimental setup to investigate the impact of modal noise on the spectra. The optical diagram of the EXOhSPEC spectrograph is also presented inside the box. The figure depicts the experimental setup where the galvanometer was implemented to mitigate the modal noise. Two identical fibres (Fibre 1 and Fibre 2) were in use. Fibre 1 was coupled between the Tungsten lamp (a white light source) and the input of the galvanometer. The galvanometer randomizes the incoming modes and collimates light onto Fibre 2. Fibre 2 was from the output of the galvanometer to the spectrograph. Spectrograph is in use to obtain high-resolution spectra of white light. This setup was termed the `GAL ON' condition. In the `GAL OFF' condition, Fibre 2 was used to feed the spectrograph directly with the white light but without the galvanometer.}
	\label{Fig:Experimental_Setup}
\end{figure*}
Measuring radial velocity (RV) is one of the key techniques to detect and characterise extrasolar planets. To measure RV accurately, high-resolution spectrographs (for instance, HARPS: \citealt{2003Msngr.114...20M}, HPF: \citealt{2012SPIE.8446E..1SM}, ESPRESSO: \citealt{2012SPIE.8446E..1RM}, CARMENES: \citealt{2012SPIE.8446E..0RQ}, PARAS: \citealt{2014PASP..126..133C}, SPIRou: \citealt{2014SPIE.9147E..15A}, NIRPS: \citealt{2017Msngr.169...21B}, NEID: \citealt{2018AAS...23124608A}, and MAROON-X: \citealt{2018SPIE10702E..6DS}) have been deployed on the different telescopes. However, the current generation of RV instruments cannot achieve the required precision (10 cm/s) to detect earth-like exoplanets orbiting sun-like (G-type) stars \citep{2016PASP..128f6001F}. To achieve such precision, it is important to maintain prolonged and extreme stability and accurate wavelength calibration of the instrument. A significant obstacle lies in stabilizing the instrument's illumination, particularly as its profile changes with varying illumination conditions \citep{1992PASP..104.1244H}.


As a remedy, the current generation of seeing-limited RV spectrographs use multimode (MM) fibres, in general, to feed the light to the spectrograph from the telescope \citep{Hill:80, {1988ASPC....3...63A}}.
In addition, utilizing the fibre effectively decouples the physical link between the instrument and  telescope's focal point, resulting in a substantial improvement in detection sensitivity. A large core MM fibre is generally used to maximise the light gathered.
However, due to the finite modes in MM fibres and the potential for subtle illumination changes, there is a concern about signal distortion by the interference of the modes as they propagate along the fibre, leading to the characteristic speckle associated with modal noise \citep{Mickelson:83}. Modal noise, characterised by this speckle pattern, arises from the interference of the finite modes in the fibres.
The amount of distortion depends on the wavelength of light, core diameter, refractive index, state of the fibre and fibre coupling and is also sensitive to environmental variation such as temperature \citep{2018ApJ...853..181P}. Furthermore, the term modal noise was introduced as the effect that distorts the signal and appears as amplitude modulation \citep{Epworth:1978}. When using a fibre to feed a spectrograph, modal noise can also be observed as Focal Ratio Degradation (FRD) \citep{Ramsey:1988}. Recent studies found that suitable tapered graded-index (GI) fibres provide an alternative way for efficient focal ratio reduction in fibre-fed spectrographs \citep{CHOOCHALERM2021102632, CHOOCHALERM2023103140}. \citet{FLORIS2020102116} showed that modal noise in GI MM fibres can be reproduced by geometrical-optics-based models. The role of modal noise on the spectrograph performance was extensively investigated in \citet{2019A&A...632A..21O} as it depends on the illumination of the source and in some cases the modal noise is negligible. Nonetheless, the modal noise has an adverse impact on the precision of RV measurements when left unmitigated \citep{2014ApJ...786...18M}. 

Over the years, various techniques have been proposed for mitigating modal noise in MM fibres, including single mode scramblers \citep{2012SPIE.8446E..8ZS, 2013PASP..125..511S}, double mode scramblers \citep{1992PASP..104.1244H, 2013PASP..125..511S, 2014SPIE.9147E..6BR, 2015PASP..127.1027S, 2015ApJ...806...61H, 2019A&A...632A..21O}, rotating double scramblers \citep{2020SPIE11203E..1OR}, dynamical diffusers with integrating sphere and rotating mirror \citep{2014ApJ...786...18M}, hand agitation \citep{2012SPIE.8446E..8JM, 2014ApJ...786...18M} mechanical agitators \citep{2012SPIE.8446E..8JM, 2018ApJ...853..181P, 2019A&A...632A..21O} and photonic lanterns \citep{2005OptL...30.2545L, 2015AdOP....7..107B, Yerolatsitis:17, 2020MNRAS.497.3713P}. 
Nevertheless, the suppression of modal noise in MM fibres continues to pose a challenge, particularly for fibres with relatively small core diameters.

Due to modal noise issues in MM fibres, single-mode fibres have been proposed as an alternative for next-generation RV spectrographs \citep{2016SPIE.9908E..7HS, 2014Sci...346..809C, 2015AAS...22525732B, 2016SPIE.9908E..19C}. 
Light in single-mode fibres propagates in a single spatial mode. This not only eliminates the modal noise issue encountered when using MM fibres but also enhances the stability of the output illumination. Nonetheless, the performance of single-mode fibre spectrographs is limited \citep{2015ApJ...814L..22H}, due to the narrow operational bandwidth in the optical range (typically 100 - 200 nm in the visible, \citealt{2018ApJ...853..181P}) and the interaction between the fundamental polarization modes \citep{Monerie1980}.

In this paper, we describe an off-the-shelf, easy-to-use technique to mitigate modal noise for small core (10 $\mu$m) MM fibres. We have investigated the effectiveness of our solution on a high-resolution spectrograph, EXO-planet high-resolution SPECtrograph (EXOhSPEC; \citealp{2019SPIE11117E..0ZL, 2019SPIE11116E..1GK, 2021PASP..133b5001J}) using step-index (SI), GI and tapered GI optical fibres. The paper is structured as follows: Modal noise on EXOhSPEC is presented in Section 2.  The experimental setup for our study is described in Section 3, while Section 4 covers our results and discussion. Finally, the summary of the study is presented in Section 5.
\begin{figure*}
	\centering
	\includegraphics[width=3.0in, height=2.5in]{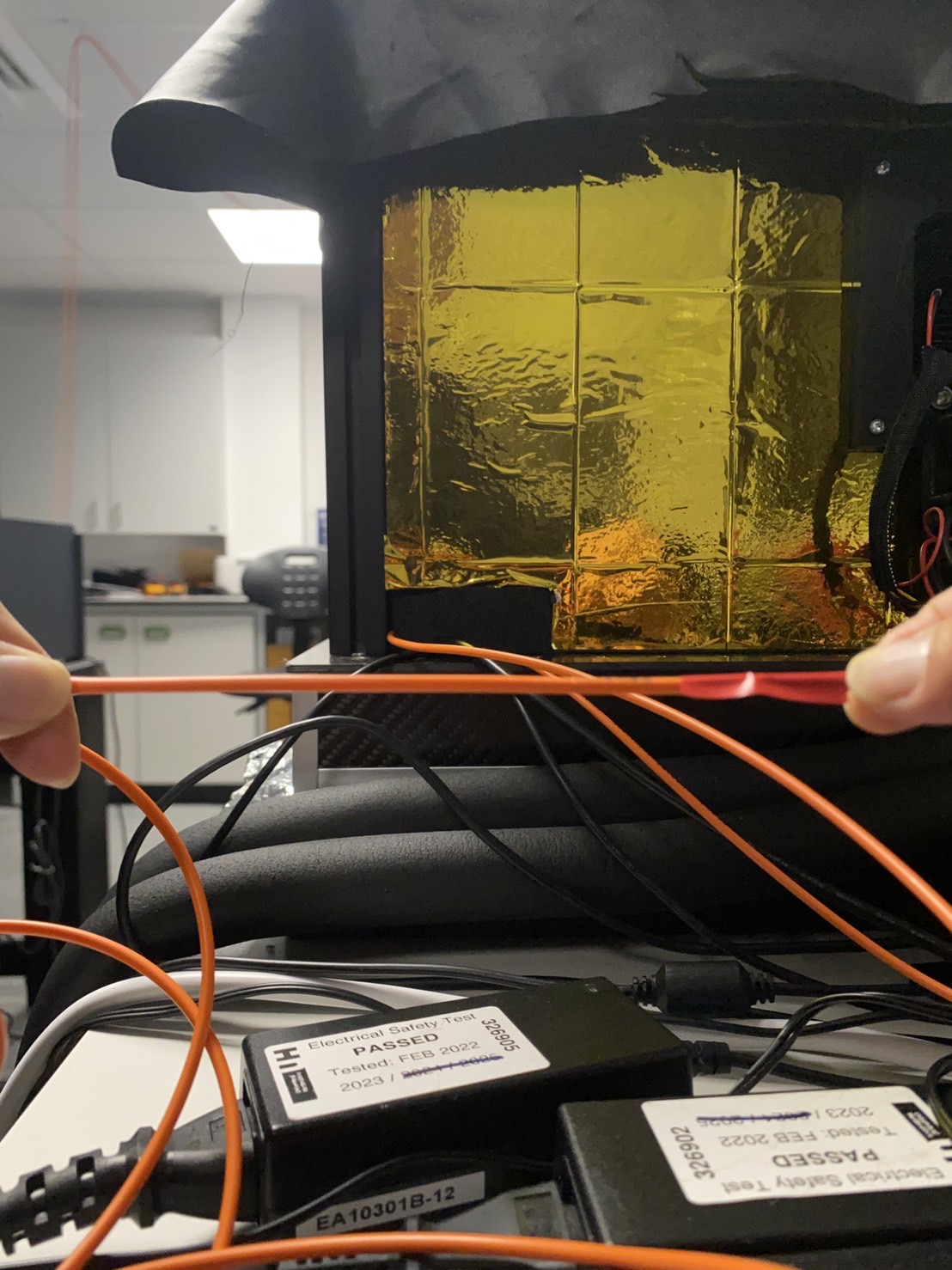}
	\includegraphics[width=3.0in, height=2.5in]{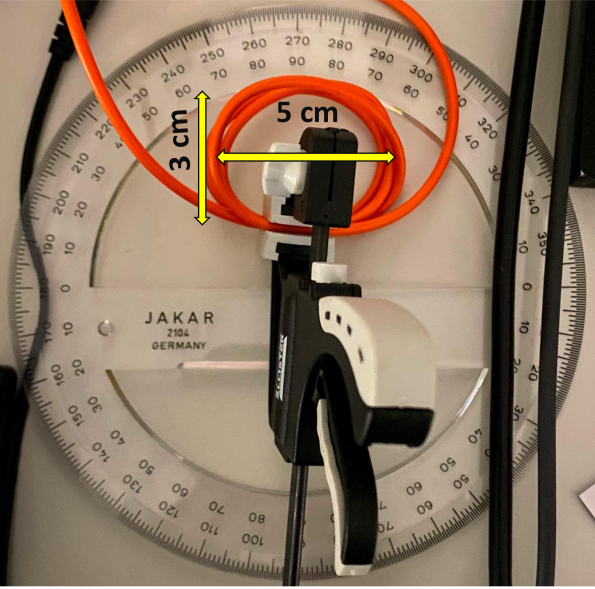}
 	\caption{Illustration of disturbing (left: stretching and right: squeezing) the fibre to study the modal noise on EXOhSPEC. The stretching and squeezing were performed by pulling the fibre in the opposite direction and by rolling the fibre with a diameter of 3 cm, respectively. A 10 $\mu$m core 0.1 numerical aperture multimode SI fibre was used for this test.}
	\label{Fig:ST_SQ_Fibre}
\end{figure*}

\section{Modal Noise on EXO\lowercase{h}SPEC with fibres}
EXOhSPEC is an actively controlled fibre-fed double-pass optical spectrograph with a spectral resolution of $>$70,000 and is designed for 10 $\mu$m core fibre with 0.10 Numerical Aperture. It was built principally from off-the-shelf components such as a collimator/camera lens, cross-dispersing prism, grating and CMOS detector and has overall dimensions of approximately 30 $\times$ 40 cm $\times$ 50 cm. The main instrument is currently under development for the 2.4 m Thai National Telescope at the National Astronomical Research Institute of Thailand (NARIT). A prototype of the instrument is at the University of Hertfordshire (UH), United Kingdom where new techniques are tested to improve the performance and stabilisation of the spectrograph. The aim of this project is to develop an efficient high-resolution spectrograph for exoplanet study as well as to reduce the size and cost with only a modest impact on the performance. 

A schematic diagram of the spectrograph is shown inside a rectangular box in Fig.~\ref{Fig:Experimental_Setup}. Additional details on the spectrograph can be found in \cite{2021PASP..133b5001J}. As mentioned the EXOhSPEC is a fibre-fed spectrograph, the fibre link is required in order to achieve the high-precision performance. However, modal noise can limit the optimisation of a fibre-feeding system for astronomical observations.

As the instrument is designed for a 10 $\mu$m core fibre, a 10 $\mu$m core Thorlabs SI multimode fibre was set up in the first place for data acquisition. The significant amount of modal noise on the UH prototype was evident in our previous study \citep{2021PASP..133b5001J}. The extracted spectrum of continuum frames showed apparently periodic features of substantial strength. The strength of wavy spectral features was found to decrease as measurements shifted from redder (900 nm) to bluer (460 nm) wavelength regimes. This is because the number of modes at the bluer end is relatively higher than the redder end and thus the modes are averaged out at the bluer end. This observation agrees well with the previous studies (see, \citealt{2001PASP..113..851B}). 

\begin{figure}
	\centering
	\includegraphics[width=3.2in, height=2.5in]{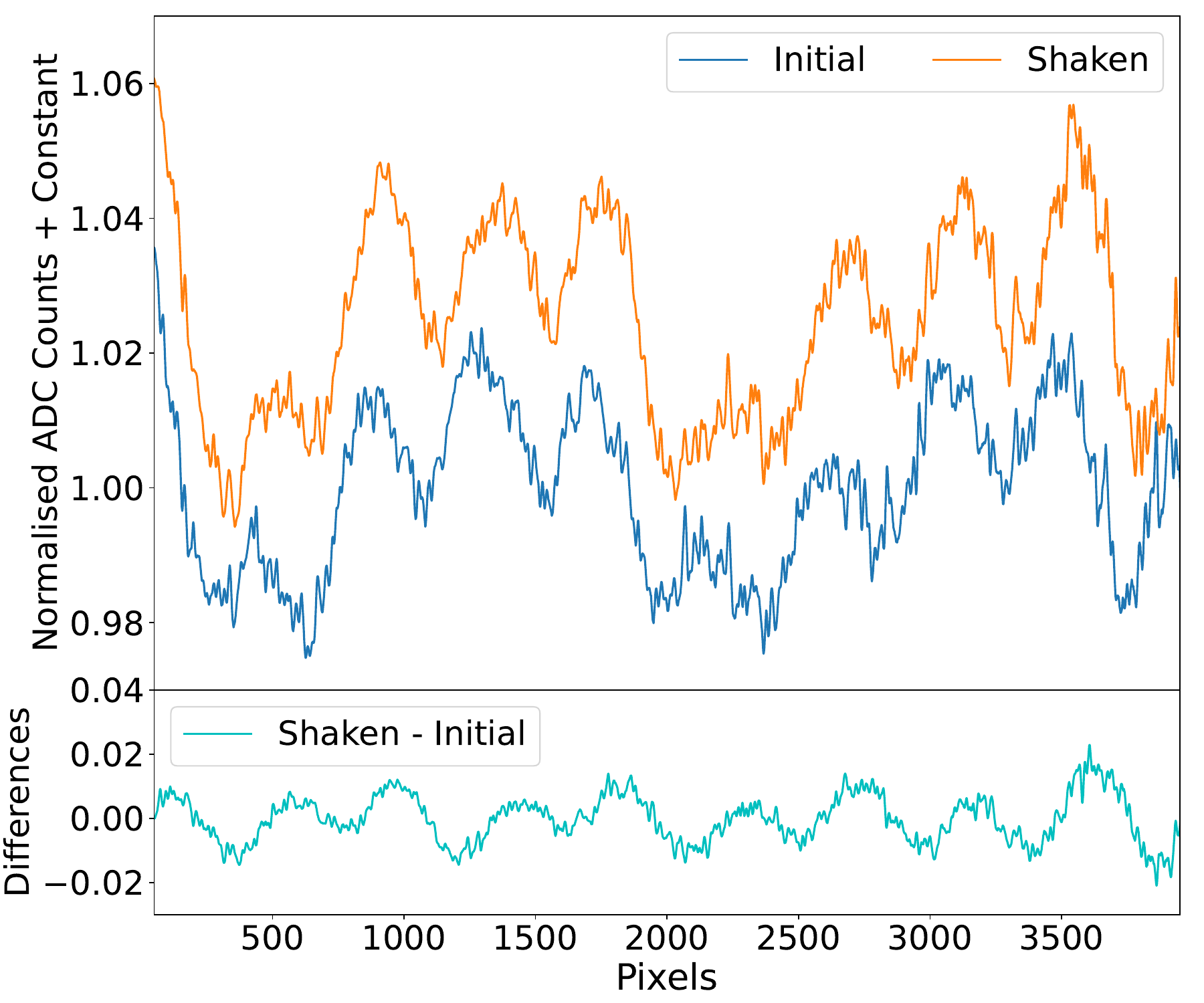}
         \includegraphics[width=3.2in, height=2.5in]{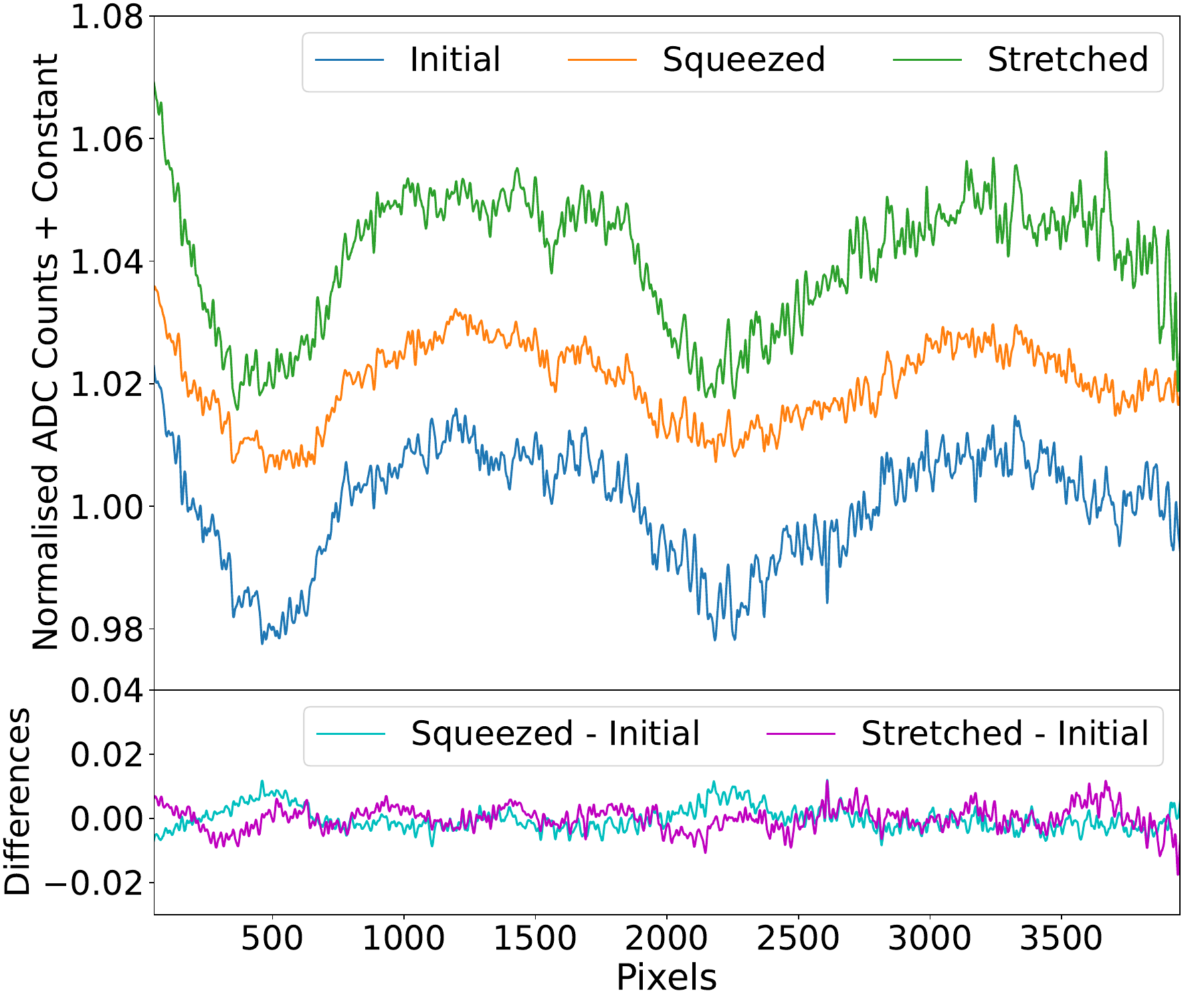} 
	\caption{Illustration of the modal noise impact on the white light spectra when a 10 $\mu$m SI fibre was shaken (upper panel) as well as squeezed and stretched (lower panel). The data for the two panels were obtained for the same fibre after setup on different dates.}
	\label{Fig:instrumental_modal_noise}
\end{figure}

A re-setup of the experiment was required to test the impact of fibre movement on modal noise. For that, the fibre was disturbed by manual shaking, stretching and squeezing as shown in Fig.~\ref{Fig:ST_SQ_Fibre}. The spectrum of the white light was obtained through the disturbed fibre. For a preliminary assessment, the spectrum of white light was extracted using MaxIm DL\footnote{\url{https://diffractionlimited.com/product/maxim-dl/}}. The average counts of a user-defined box with a height (i.e., along the dispersion direction) of 4000 pixels and a width (i.e., along the cross-dispersion direction) of 50 pixels around a certain order are adopted for the spectrum extraction. The extracted spectrum is then fitted with a 4$^{th}$ order polynomial for normalisation. The normalised spectra are displayed in Fig.~\ref{Fig:instrumental_modal_noise} for different fibre conditions. Looking at the spectra, we can see that the spectrum changes with the change in the fibre's state. In fact, the overall pattern of modes between the same fibres taken after setup on different days is larger than differences that we can induce in the fibres by shaking, squeezing or stretching. To make the speckle pattern stable and suppress this modal noise, we investigated the use of optical reimaging agitation to mix the modes using a galvanometer as demonstrated in the following section.

\section{Experimental Setup}
We investigated our galvanometer-based technique on the UH prototype of the spectrograph with fibres of different core sizes and refractive index profiles. The core size and refractive index profile of a fibre defines the number of modes, hence the amount of modal noise for a fixed illumination. Moreover, we are interested in testing both MM SI and GI fibres. SI fibres are traditionally used in the fibre feed spectrographs (for example, \citealt{2018PASP..130f5003I, 2018ApJ...853..181P}) because of its relatively high transmission in the blue region. However, with GI fibre, refocusing of the signal within the core is increased because of the parabolic refractive index profile. Also, the pulse dispersion for a given distance is smaller in GI fibre than SI fibre. In addition, we adopted conventional circular core fibres for our experiment instead of other types of fibre such as rectangular and octagonal core fibres as our main objective is to test the effectiveness our method in suppressing modal noise.

\subsection{Description of Galvanometer} \label{sec:Description_of_Galvanometer}

\begin{figure}
	\centering
	\includegraphics[scale=0.21]{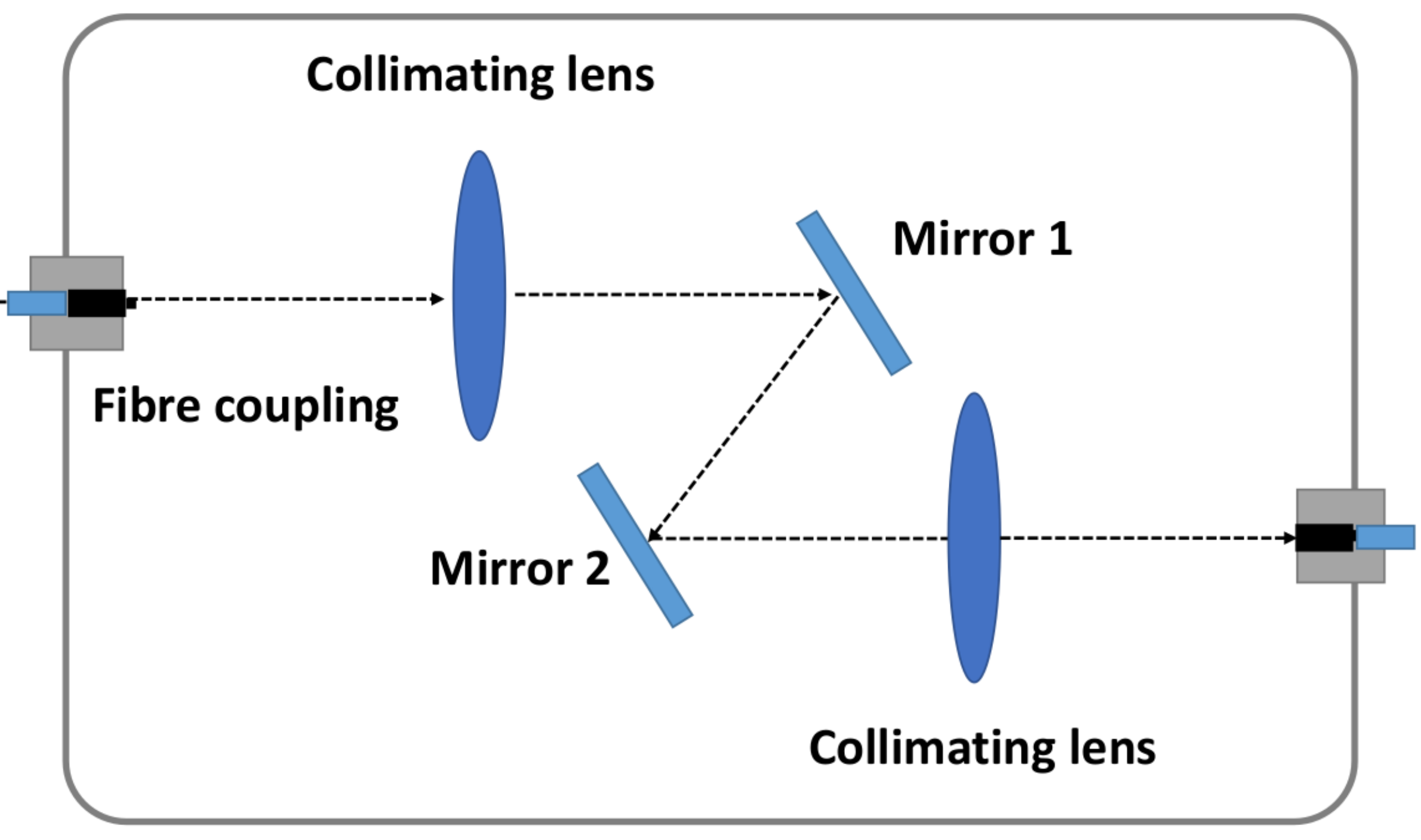}
	\caption{Schematic layout of the galvanometer used for suppressing the modal noise. Two motorized rotating flat mirrors that are used to mix the modes can be controlled by altering the amplitude and voltage of the applied signal to each channel independently through a voltage generator.}
	\label{Fig:galvanometer_operation}
\end{figure}

The galvanometer system consists of a Thorlabs GVSX02 dual-axis galvo, and two fibre collimators (TC06FC-633). There are two movable flat mirrors in the galvanometer, Mirror 1 and Mirror 2. For detailed configuration of the galvanometer, we refer to the catalog\footnote{\label{footN1}\url{https://www.thorlabs.com/catalogpages/Obsolete/2023/GVS302.pdf}}. The reflectivity of two mirrors is 95\%\footref{footN1} at 635 nm. Fig.~\ref{Fig:galvanometer_operation} presents a schematic of the galvanometer setup. One end of Fibre 1 is connected to a white light source. The other end is connected to the input fibre coupler of the galvanometer. Light from the input fibre, Fibre 1, is collimated by the first collimator, directed to the first movable mirror Mirror 1, and reflected towards the second movable mirror Mirror 2. The reflected light from the Mirror 2 is refocused to the input of Fibre 2 by the output fibre collimator. Fibre 2 is connected to the spectrograph.

Both mirrors are mounted on motorized rotating mounts. Using this configuration we are able to move the two mirrors independently. The mode pattern from Fibre 1 is moved in X and Y and is collimated to focus on Fibre 2. The modes of Fibre 2 are excited randomly and hence, the output illumination pattern varies continuously throughout an appropriately long integration time. As a result, the speckle pattern is averaged out efficiently at the exit of Fibre 2. Thus, two movable mirrors mix the modes before recoupling the light from the input to the output fibre.

The movement of the mirrors can be controlled by applying sinusoidal signals on two channels (CH 1 and CH 2) of a voltage generator. Controlling the frequency and amplitude of the signal, we can control the movement of mirrors, hence the mixing of modes. CH 1 controls the Y-axis movement of the Mirror 1 and CH 2 controls the X-axis movement of the Mirror 2.

The optimum amplitudes and frequencies for operating the galvanometer were explored and this is discussed in Sec~\ref{sec:Results_and_Discussion}. Before setting up the galvanometer the two mirrors were optimally aligned using a fibre coupled laser by applying a small DC offset voltage through the voltage generator to optimise the output for a given input to the galvanometer. We set the frequency and amplitude of the signal to each of the mirrors at 1 Hz and 1 mV, respectively, and changed the DC offset voltage to find optimum settings of the galvanometer output using a power meter. The DC offset voltages for the maximum throughput power were found to be $-$65 mV and 1.48 V for CH 1 and CH 2, respectively. We repeated the test for different values of frequencies and amplitudes and found that the DC offset voltages remained the same. This indicates that there is no dependence of the applied frequencies and amplitudes on the DC offset voltage. Once the DC offset voltage is fixed at the beginning of the experiment, it remains the same during the experiment. Furthermore, we have noticed no notable change in DC offset value in our day-to-day experiments.

We determined the angle of rotation of two mirrors due to the applied signals to the voltage generator by removing the collimators. We calculated the angle of rotation of the two mirrors in response to the signals applied to the voltage generator. The rotation of the mirrors due to applied voltage at CH 1 and CH 2 to be 36.6 microradian ($\mu$rad) per millivolt (mV) and 51.8 $\mu$rad/mV, respectively. Translating the rotation of the mirrors into the movement on the output fibre, we obtained 0.5 $\mu$m and 1.3 $\mu$m per mV along X and Y-axes, respectively. For the application of 1 mV to each of the mirrors, the focal spot of the output beam would typically move $\sim$1.4 $\mu$m.

\subsection{Data Acquisition} \label{sec:data_acquisition}

\begin{table}
	 \centering
	\caption{Details on the adopted fibres for investigation.}
	\label{tab:FibreDetailed}
         \resizebox{0.46\textwidth}{!}{%
	   \begin{tabular}{lcccll}
          \hline
		Fibre & Core & Numerical & Number of & Manufacturer & Part \\
            type & Size & aperture & modes$\dagger$  & {} & number\\
		\hline
           SI50 & 50 $\mu$m & 0.22 & 988 & Thorlabs & FG050LGA \\
           GI50 & 50 $\mu$m & 0.20 & 816 & Thorlabs & GIF50E \\
           SI10 & 10 $\mu$m & 0.1 & 8 & Thorlabs & FG010LDA \\
           GI50t & 10 $\mu$m & 0.2 & 16 & University of Bath$\ddagger$ & Custom  \\
		\hline
	   \end{tabular}}
Notes: $\dagger$The length of each fibre is 5 m. The number of modes for each fibre was estimated using equation (1) of \citet{2018ApJ...853..181P} at 0.7 $\mu$m.\\
$\ddagger$The fibre used for tapering was from University of Bath and it was tapered by Thorlabs Vytran Automated glass processing workstation (GPX3400).
\end{table}

 We performed our experiment with four circular MM fibres.
 As mentioned above, the instrument is designed for a 10 $\mu$m core input fibre with 0.10 numerical aperture. We tested both 10 $\mu$m core SI (SI10) as well as GI fibres. A SI10 is commercially available, however, a 10 $\mu$m core GI fibre is not.  Thus, we connected a 50 $\mu$m GI fibre with a custom 5:1 tapered fibre (GI50t) i.e., waist of the tapered fibre has a 10$\mu$m core and so should behave approximately as a 10$\mu$m core GI fibre. The tapering process was performed on a Thorlabs Vytran GPX3400 Glass Processor. The tapered fibre generally converts the beam diameter of an input beam to a small diameter and rejects relatively high-order modes. We refer to \citet{CHOOCHALERM2021102632, CHOOCHALERM2023103140} for details on the transmission properties of tapered optical fibres. Additionally, we investigated commercially available 50 $\mu$m SI (SI50) and  50 $\mu$m GI fibres for comparison purposes. We did not test fibres having larger core diameters than 50 $\mu$m core as the modal noise reduces with the enlargement of fibre core size.
The additional details of the fibres are listed in Table~\ref{tab:FibreDetailed}. 
A white light source (tungsten lamp - Thorlabs SLS201L/M) was used to illuminate the fibre on test (see Fig. \ref{Fig:Experimental_Setup}). Following the procedure described below, we were able to test the performance of our method. 

\begin{itemize}
    \item GAL ON: Two identical fibres, namely Fibre 1 and Fibre 2, were used. Fibre 1 established a connection between the white light source and the input of the galvanometer. Fibre 2 connected the output of the galvanometer to the spectrograph as shown in Fig.~\ref{Fig:Experimental_Setup}. The modes of second fibre were being agitated by the galvanometer.

    \item GAL OFF: No galvanometer was inserted into the light path. Fibre 1 is removed and Fibre 2 was connected from the white light source directly to the spectrograph.  There was no agitation in this case.
\end{itemize}
The goal of our two measurement setups was to compare the modal noise between sets of spectra obtained both with and without agitation and thereby verify the effectiveness of our method. The typical exposures were 0.4 s for 50 $\mu$m fibres (SI50 and GI50) and 5 s and 10 s for GI50t and SI10, respectively.

\begin{figure*}
	\centering
	\includegraphics[width=1.65in, height=1.5in]{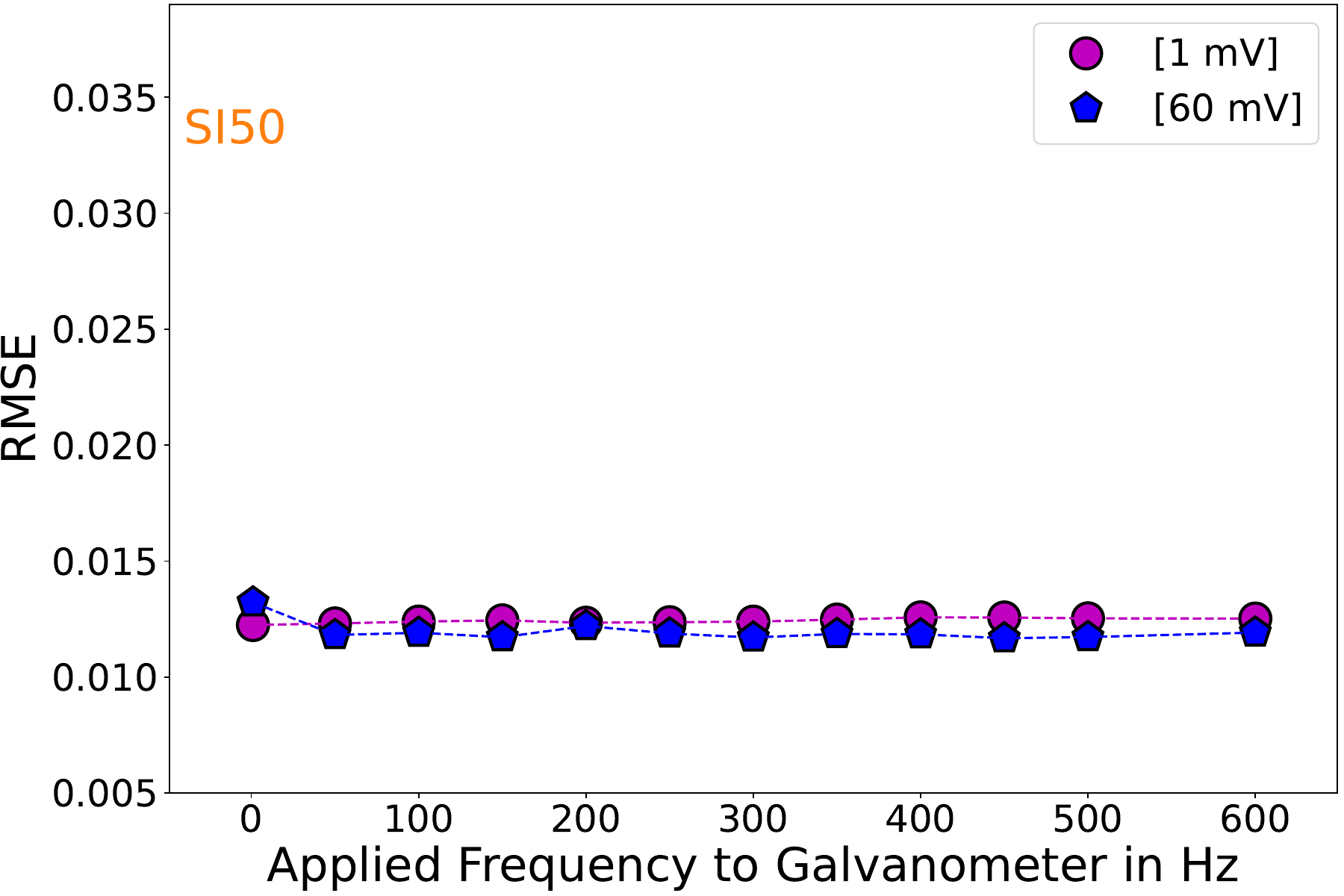}
	\includegraphics[width=1.65in, height=1.5in]{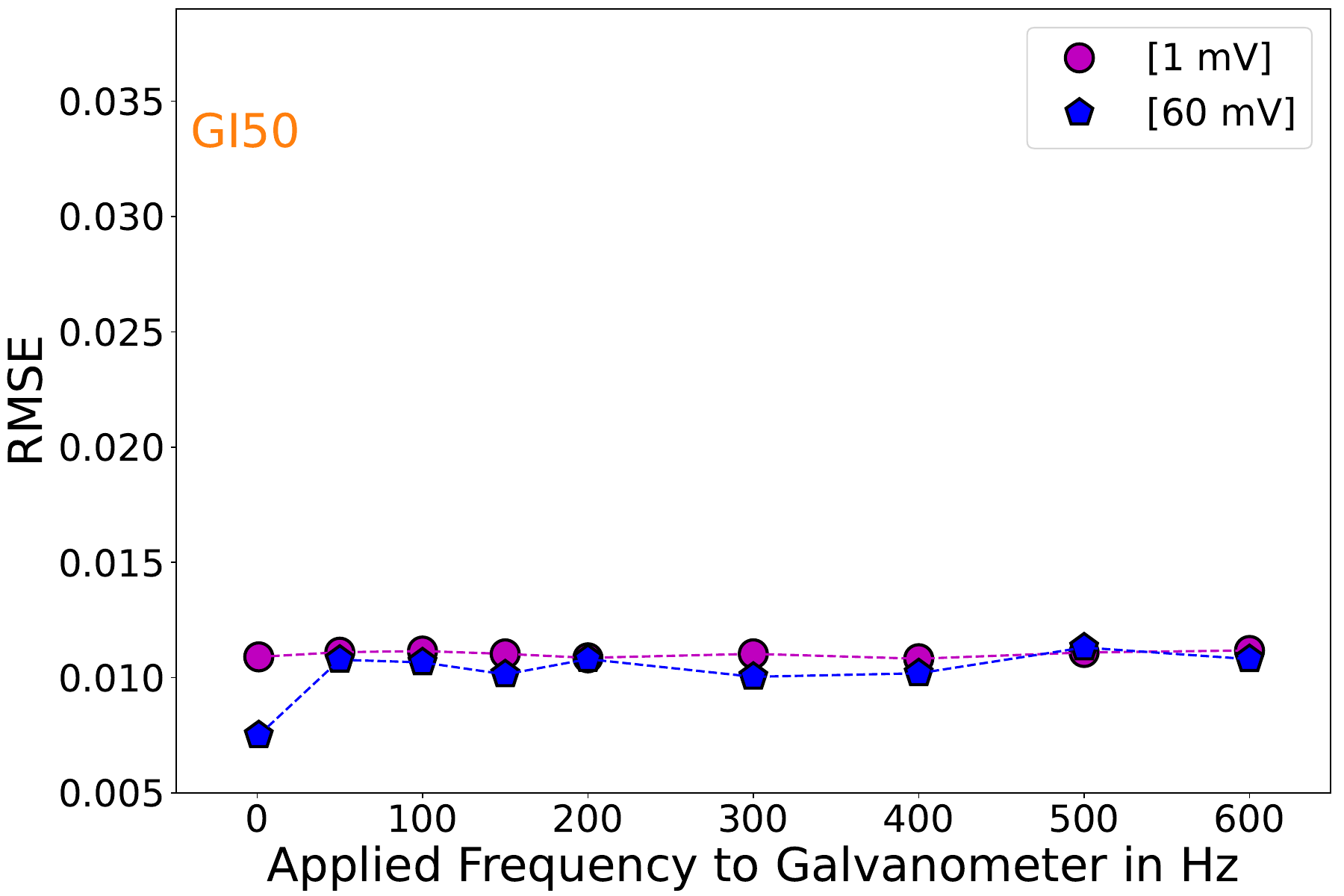}
	\includegraphics[width=1.65in, height=1.5in]{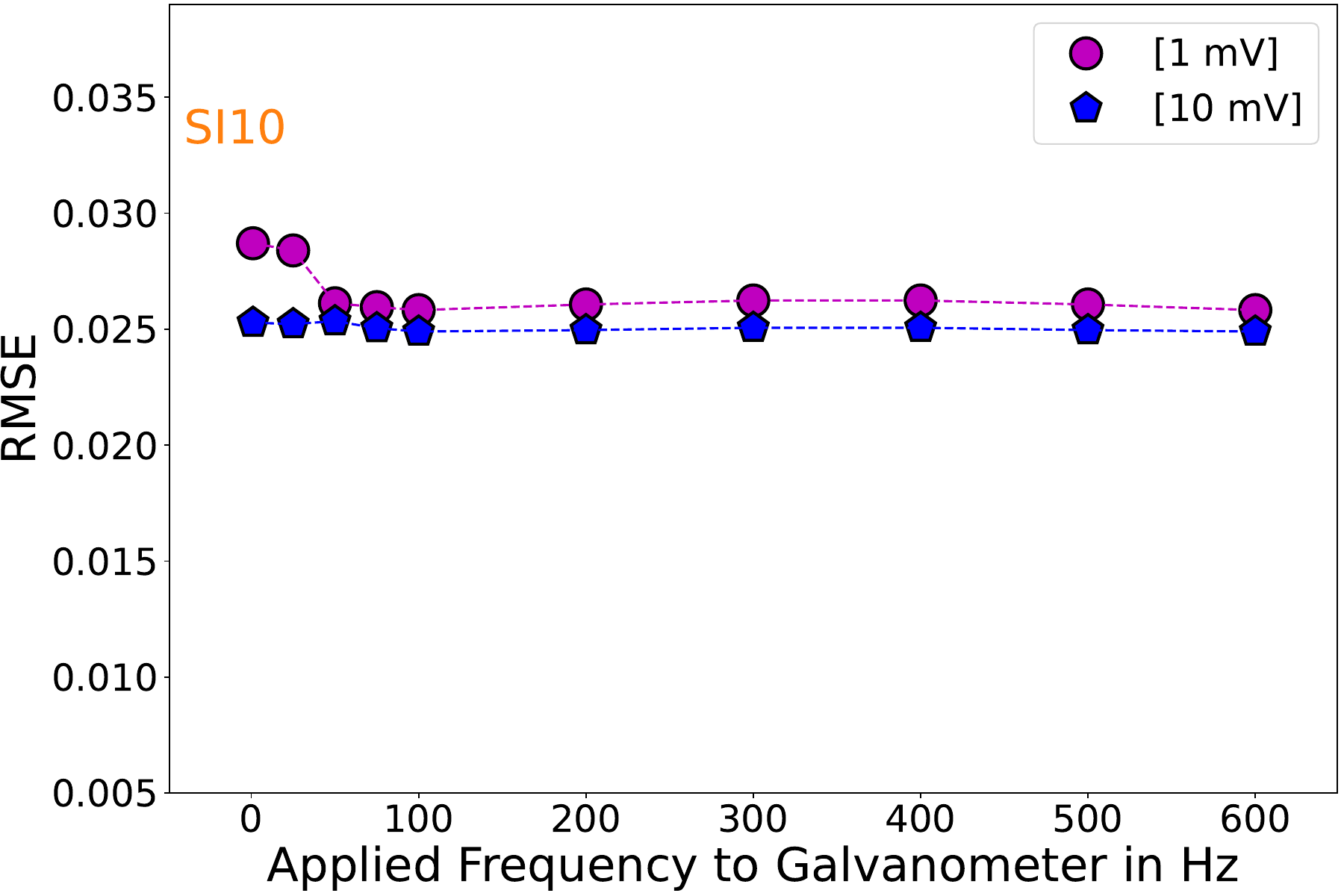}
	\includegraphics[width=1.65in, height=1.5in]{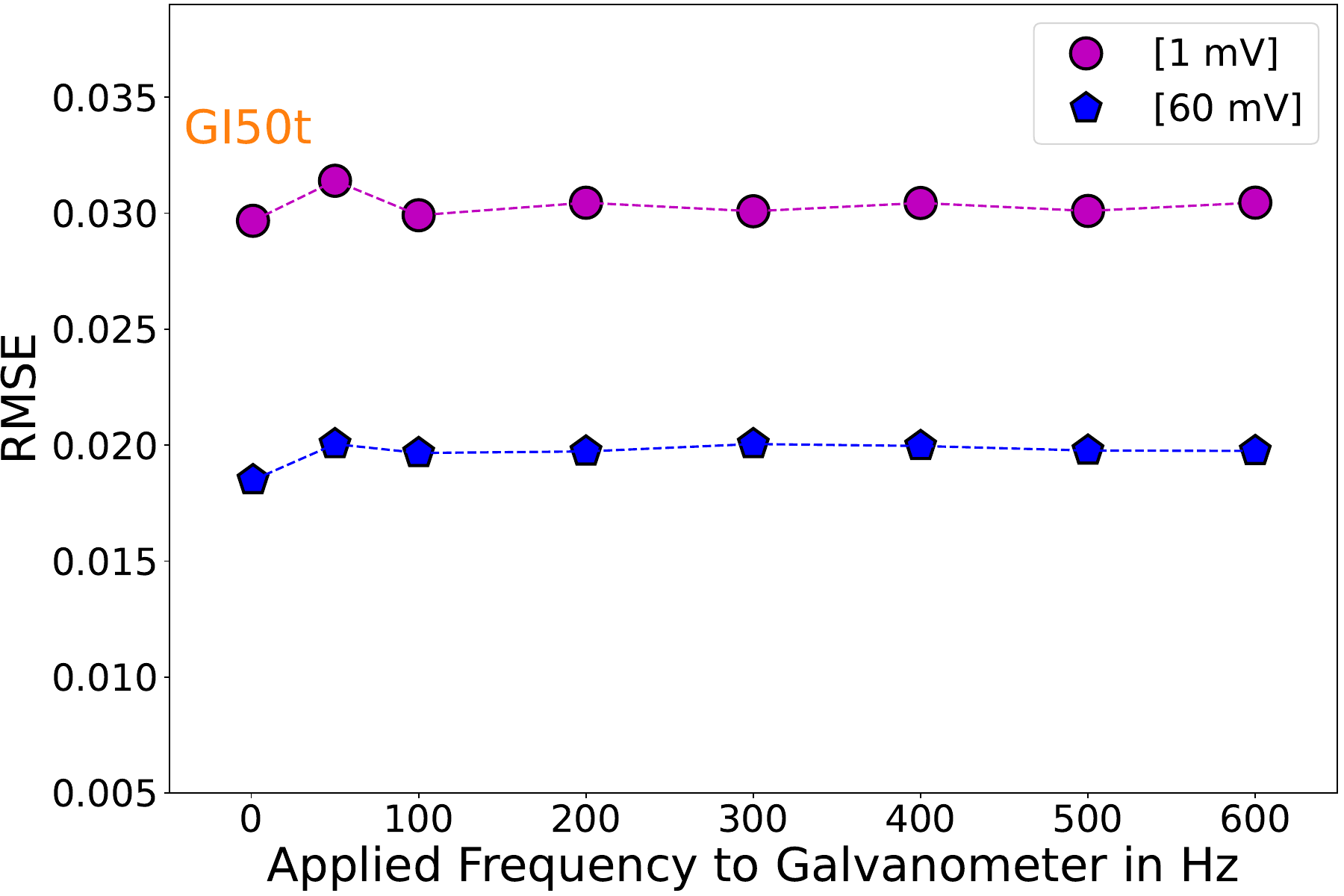}
	\caption{RMSE variation with applied frequencies to the galvanometer for different fibres (SI50, GI50, SI10, and GI50t, respectively) to find the optimum frequency for operating the galvanometer. We carried out our investigation for two amplitudes, 1 mV and 60 mV. The violet symbols correspond to 1 mV. The blue symbols represent 60 mV for SI50, GI50, and GI50t and 10 mV for SI10 fibre. All plots are made on the same scale for relative comparison.}
	\label{Fig:rmse_frequency}
\end{figure*}

\begin{figure*}
	\centering
	\includegraphics[width=1.73in, height=1.47in]{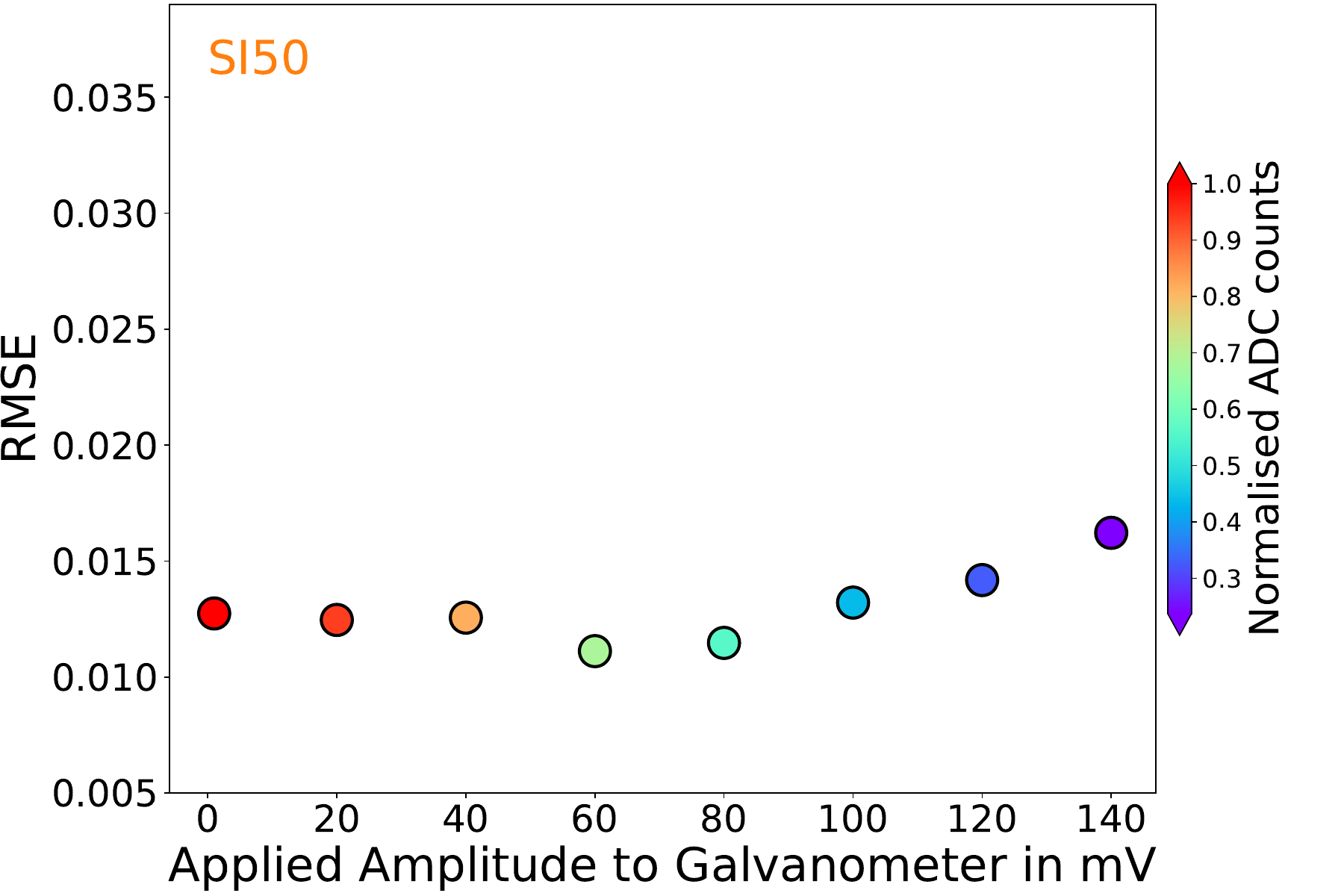}
	\includegraphics[width=1.73in, height=1.47in]{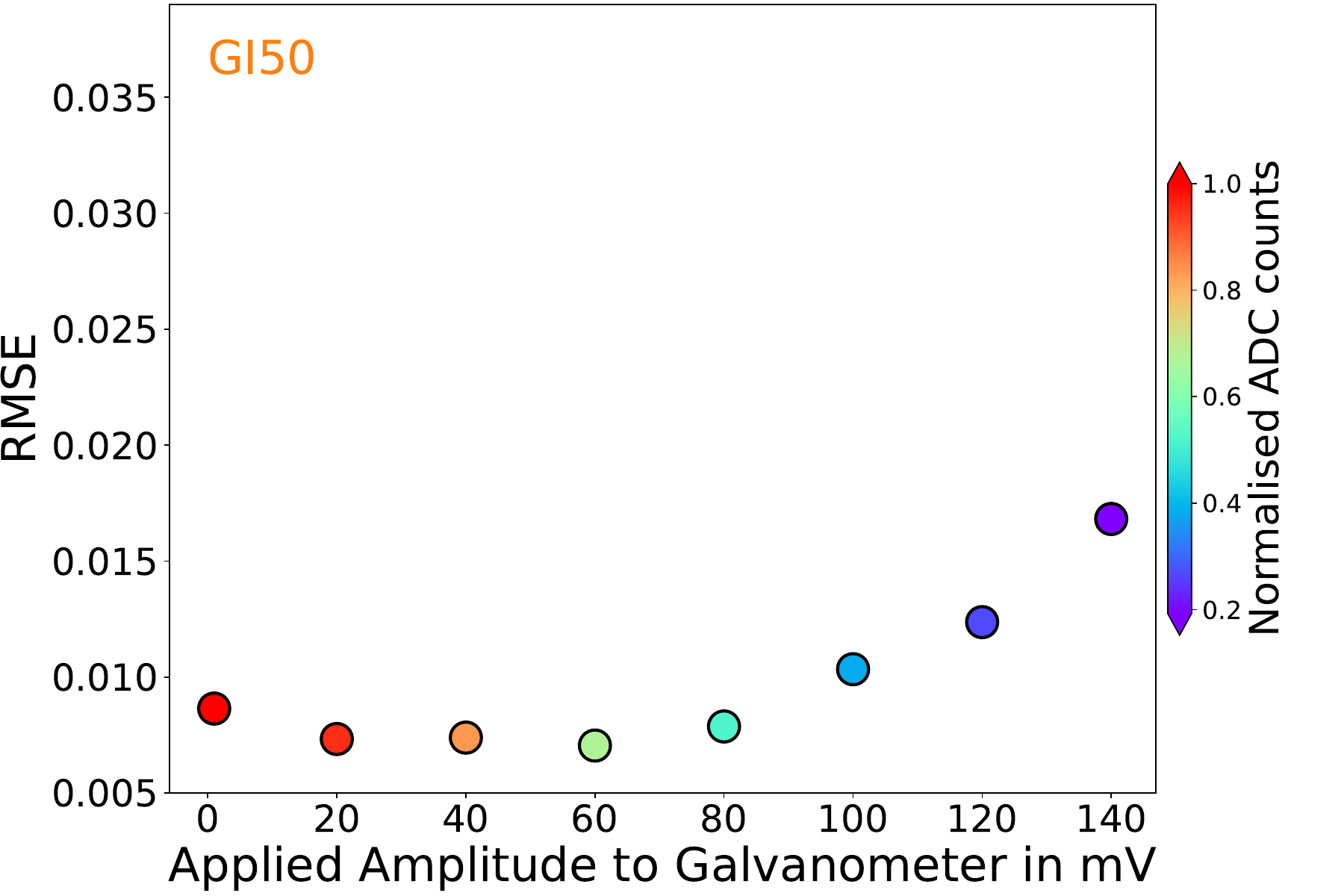}
	\includegraphics[width=1.73in, height=1.47in]{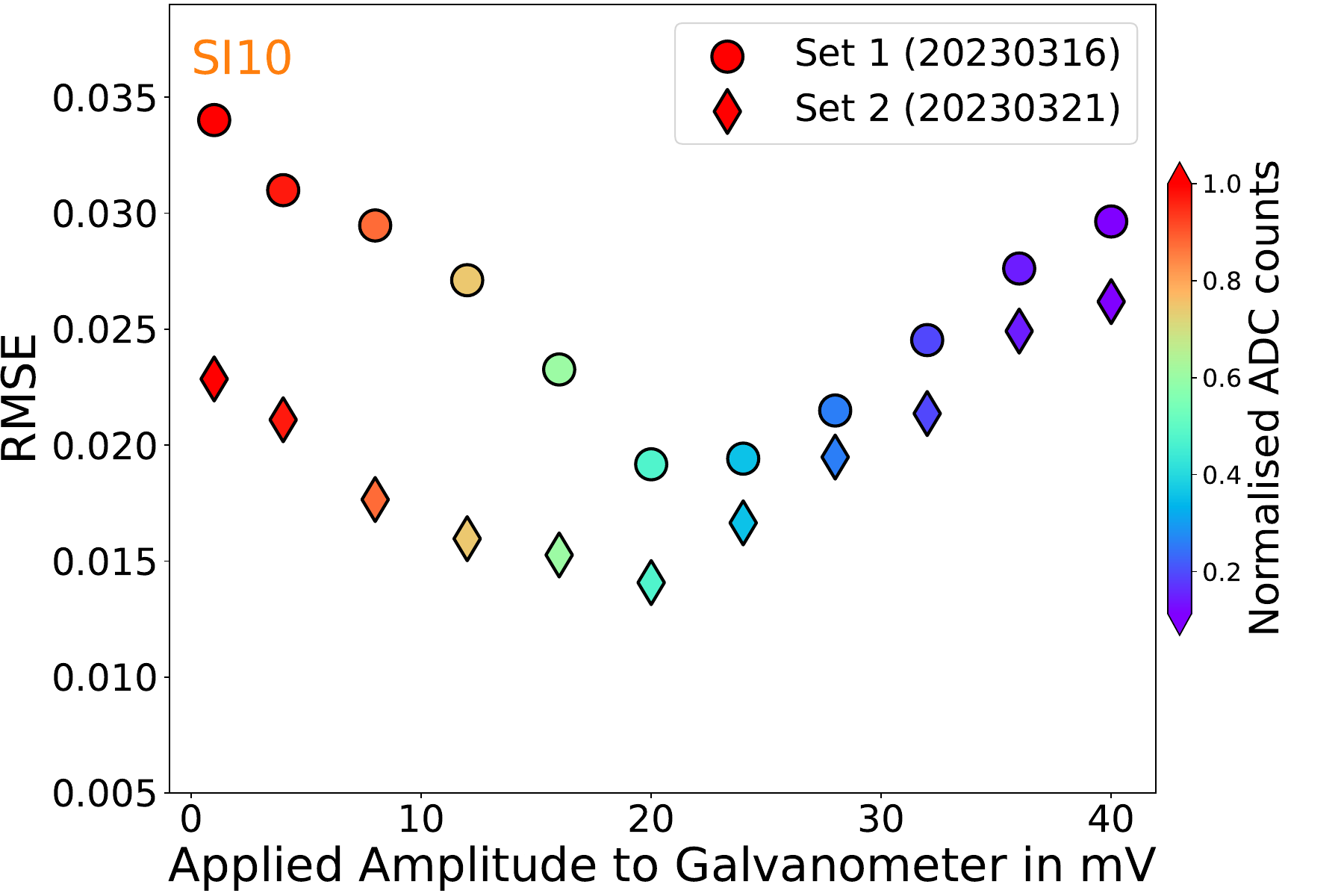}
	\includegraphics[width=1.73in, height=1.47in]{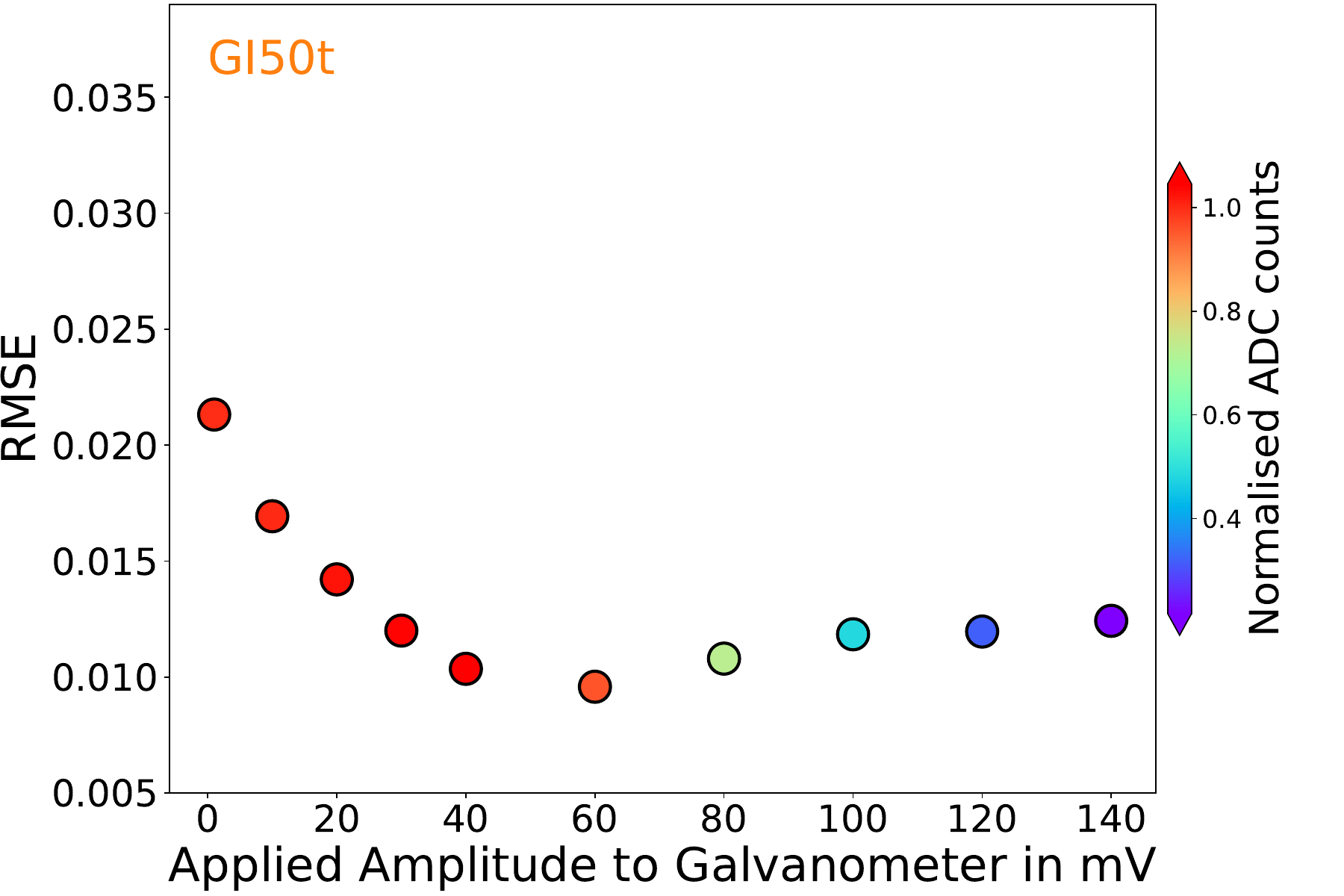}
	\caption{RMSE variation with applied amplitudes to the galvanometer for different fibres (SI50, GI50, SI10, and GI50t, respectively) to find the preferred amplitude for operating the galvanometer.  The applied frequencies were 89 Hz and 97 Hz to CH 1 and CH 2 of the frequency generator, respectively. The colour bar represents the normalised ADC counts corresponding to each RMSE measurement. See the text for details. All plots are made on the same scale for relative comparison.}
	\label{Fig:rmse_amplitude}
\end{figure*}
\begin{figure*}
	\centering
	\includegraphics[width=3.4in, height=2.75in]{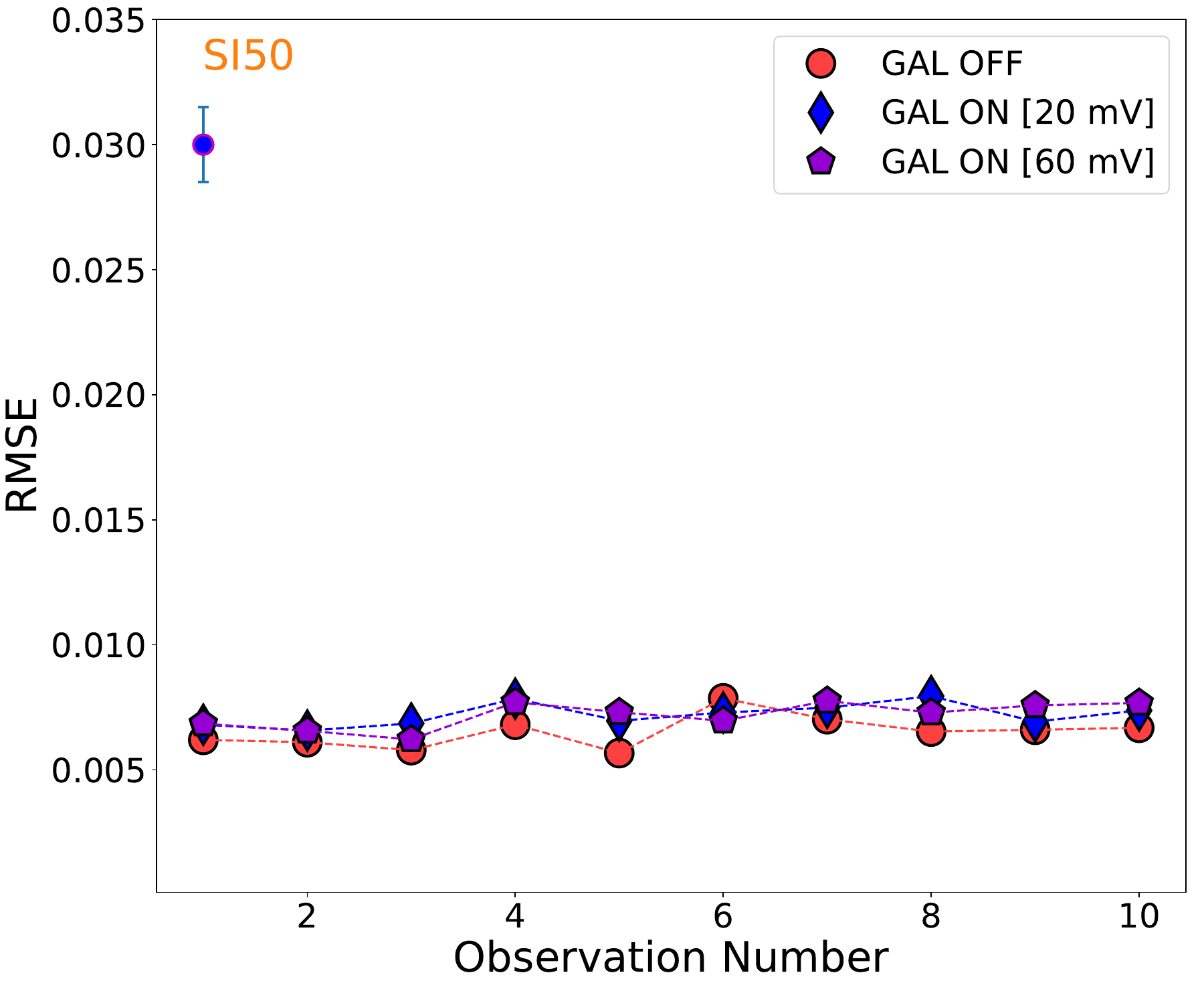}
	\includegraphics[width=3.4in, height=2.75in]{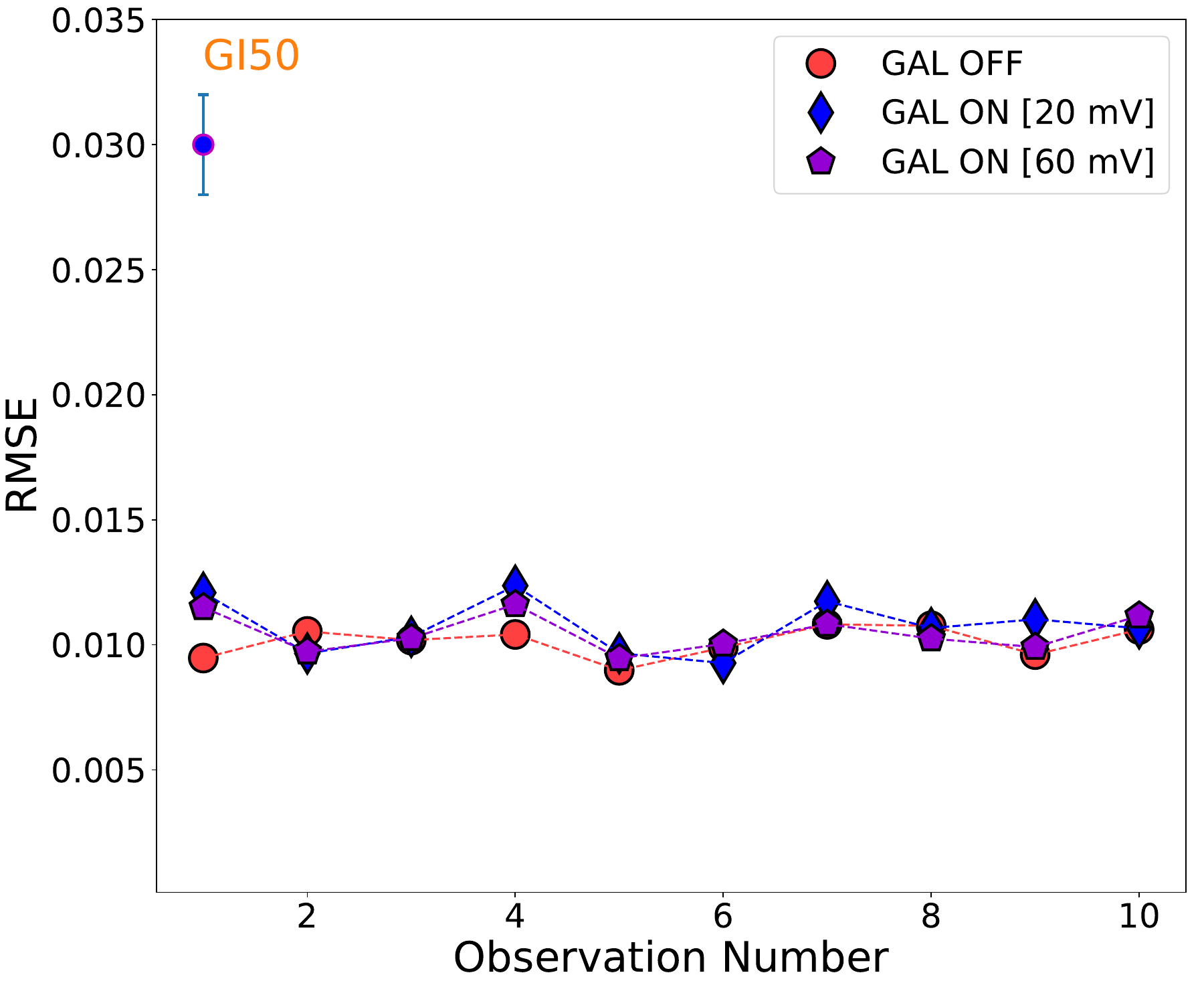}
	\includegraphics[width=3.4in, height=2.75in]{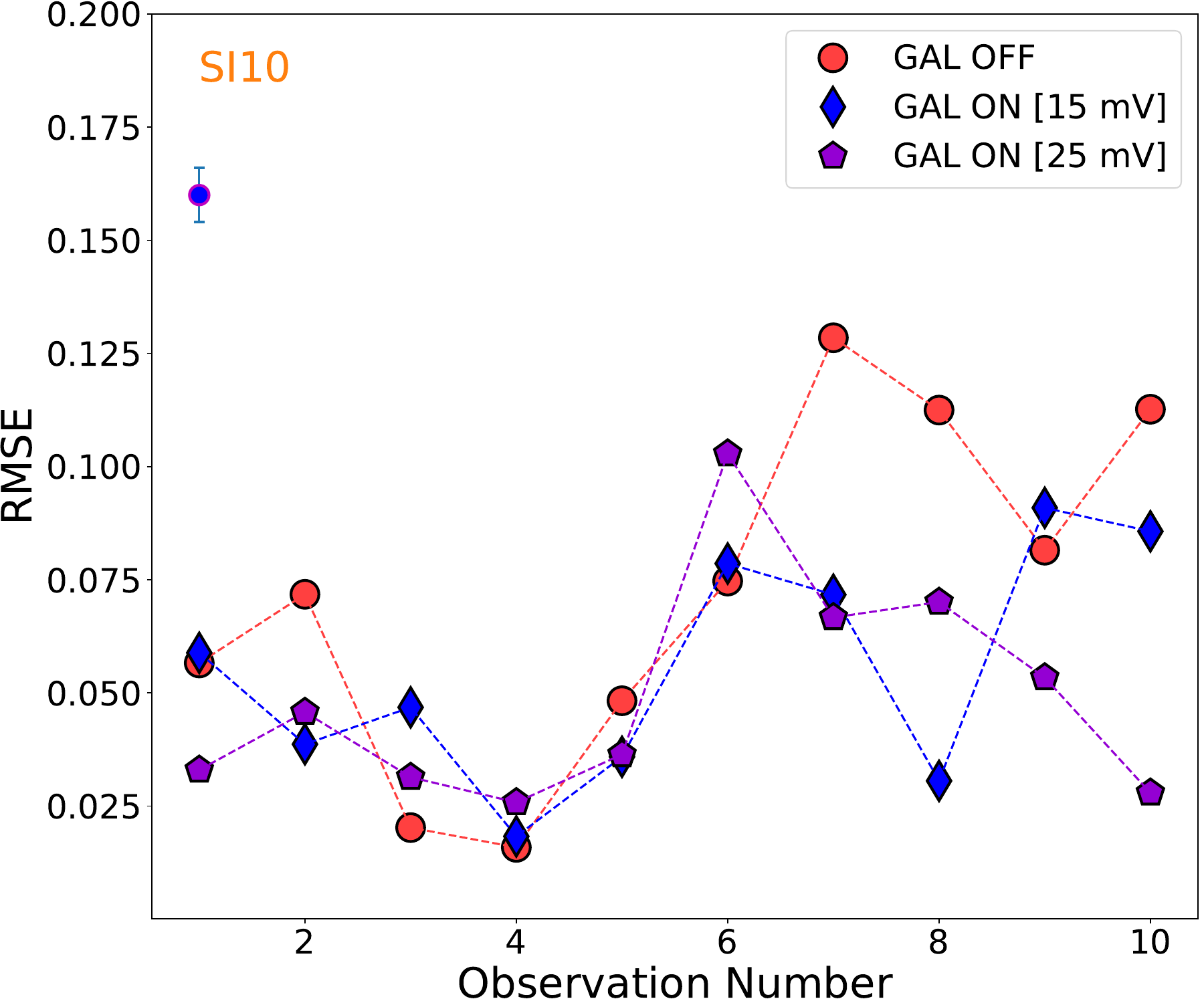}
	\includegraphics[width=3.4in, height=2.75in]{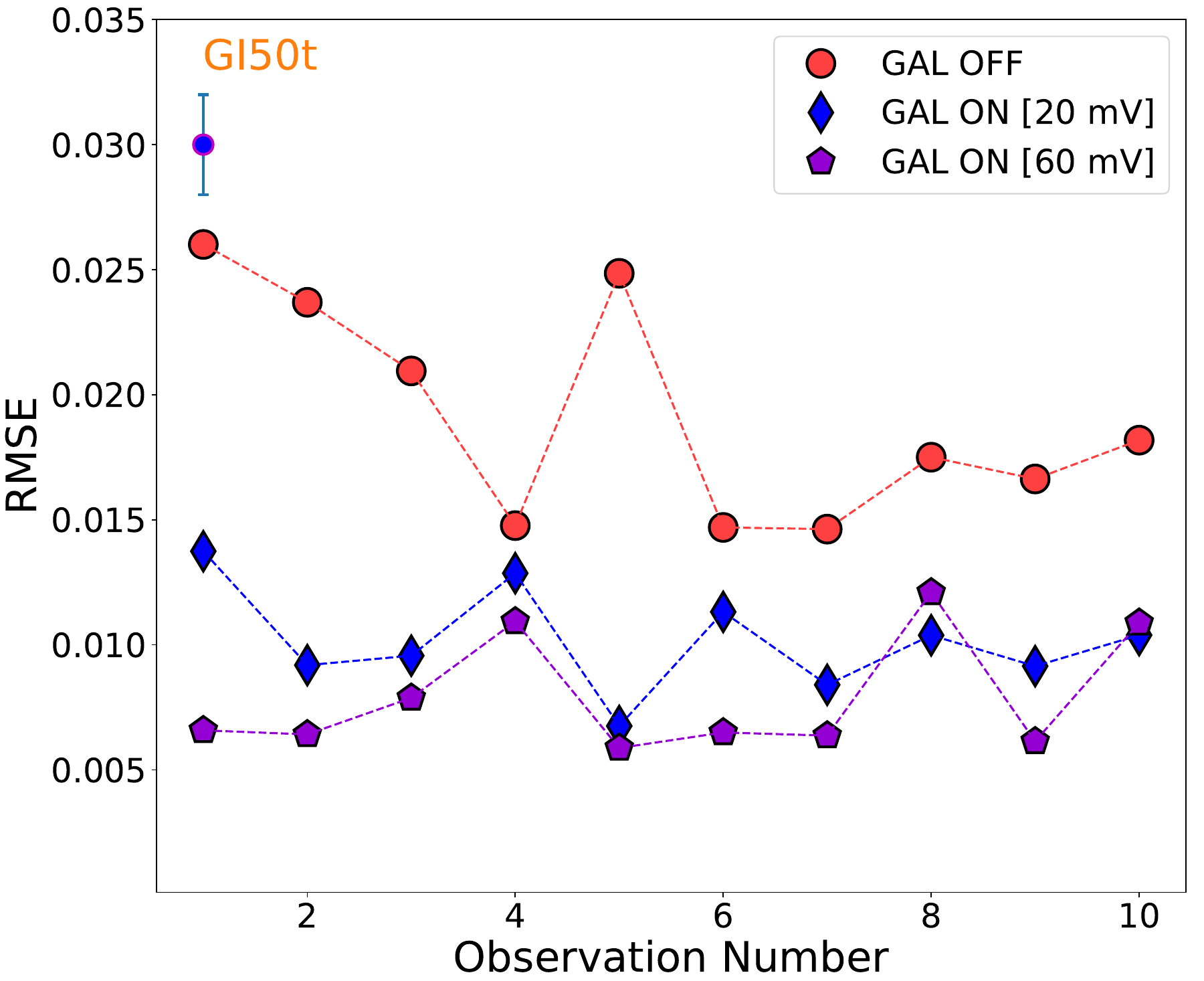}
	\caption{RMSE scatter with and without galvanometer for different fibres (SI50, GI50, SI10, and GI50t, respectively). The red symbols represent the `GAL OFF' condition. The other symbols correspond to the `GAL ON' condition. Blue error bar at the top of each plot represents typical RMSE measurement error. We carried out our experiments for amplitudes, 20 mV (blue symbols) and 60 mV (violet symbols) for SI50, GI50, and GI50t and 15 mV (blue symbols)and 25 mV (violet symbols) for SI10. The applied frequencies were 89 Hz and 97 Hz to CH 1 and CH 2 of the frequency generator, respectively. We repeated the experiments multiple times to confirm our results.  In each observation, the position of the fibre was altered to make sure that the state of the fibre (mode pattern) was not the same, and then we waited for 5 mins before taking the data. GI50, SI50 and GI50t are made on the same scale for relative comparison.}
	\label{Fig:rmse_obserN}
\end{figure*}
\subsection{Spectrum extraction} \label{Sec:Spectrum_extraction}

We performed data analysis in Python, correcting the raw dataset of the white light (image frames) for bias, dark and flat before extracting its spectra. We utilised five bias (dark) frames for bias (dark) correction, respectively. We median combined the images on a pixel-by-pixel basis and subtracted them from the image frames. We obtained five frames of white light for flat correction at the start of each day's data acquisition. After bias and dark corrections, we median combined these frames on a pixel-by-pixel basis. We smoothed the combined flat frame by implementing a median filter. We then divided it by its smoothed version for normalisation. We used the normalised flat frame for flat correction. Finally, we used the flat-corrected frame for spectrum extraction. We extracted the spectrum of a particular order, which was at the redder side of the detector, using the SpectrumExtractor tool\footnote{\url{https://github.com/indiajoe/SpectrumExtractor}}. The central wavelength of our selected order was approximately 700 nm. We picked up the order from the detector's redder side because of the relatively high throughput at this wavelength range. In addition, as the redder side propagates a relatively small number of modes, the impact of modal noise is higher on the redder side than the bluer. The same spectral order was used for all the investigated fibres for comparison purposes. A 4$^{th}$ order polynomial was found to be suitable to trace out the dispersion axis without over-fitting the data. An aperture window of ($-$6, 6) pixels was specified for spectrum extraction via sum extraction.

\subsection{Measurement of Modal Noise} \label{sec:RMSE_measurement}
To quantify the modal noise, various methods, for instance measuring the centroid drift and radial velocity error \citep{2014ApJ...786...18M, 2015ApJ...806...61H, 2018ApJ...853..181P}, signal to noise ratio (SNR, \citealt{2019A&A...632A..21O}) and barycentre precision of spectral features \citep{2020MNRAS.497.3713P}  have been used. We used root mean square error (RMSE) to measure the modal noise. The extracted continuum lamp spectrum was divided by the fitted 4$^{th}$ order polynomial described above for spectrum normalisation. The deviation of the normalised spectrum from the ideal flat normalised continuum spectrum was measured by RMSE measurement using the following equation.
\begin{equation}
\text{RMSE}(y, \hat{y}) = \sqrt{\frac{\sum_{i=0}^{N - 1} (y_i - \hat{y}_i)^2}{N}}
\end{equation}
where $y_i$ and $\hat{y_i}$ are the ideal and observed normalised flat values for $i$-th datapoint, respectively and $N$ is the total number of datapoints. The amount of modal noise is proportional to the RMSE value. The limitation of using RMSE as a proxy for modal noise quantification is the SNR of the extracted spectrum. Thus, for our analysis, we used spectra from 100 pixels to 5000 pixels of an order because the SNR is higher and more uniform.

\section{Results and Discussion} \label{sec:Results_and_Discussion}

\subsection{Frequency of Galvanometer Agitation} \label{sec:frequency_measurement}
In order to efficiently use the galvanometer the optimum frequency of operation was first evaluated. For that, the frequency of the voltage generator was varied from 1 Hz to 600 Hz keeping the amplitude fixed at a certain value for all fibres investigated as shown in Fig.~\ref{Fig:rmse_frequency}. Based on the size of the core of the fibre, we carried out our experiment with different values of amplitude. We carried out the test at two amplitudes: 1 mV and 60 mV for SI50, GI50 and GI50t, and 1 mv and 20 mV for SI10.  We extracted the spectrum for an order in each case as mentioned in Sec~\ref{Sec:Spectrum_extraction} and evaluated RMSEs following the approach described in Sec~\ref{sec:RMSE_measurement}. We found no clear evidence of frequency dependence (between 1 Hz and 600 Hz) on the RMSE measurement for any of the fibres. Nonetheless, applied frequencies of 89 Hz to Mirror 1 (CH 1) and 97 Hz to Mirror 2 (CH 2) satisfying the condition for fibre agitation frequency proposed by \cite{2001PASP..113..851B}. It states that the agitation frequency should be much greater than 1/$\tau$, where $\tau$ is the exposure time. Our values of integration are 0.4 s to 10 s, so values of frequencies (89 Hz and 97 Hz) always exceed 1/$\tau$. Furthermore, it should be noted that we used two prime numbers to reduce the magnitude of possible resonances. Also, operating the galvanometer at higher frequencies (>300 Hz) may reduce its longevity.
\subsection{Amplitude of Galvanometer Agitation} \label{sec:amplitude_measurement}
To find the optimum amplitude, we kept the frequencies of two channels constant for all fibres and varied the amplitude using a voltage generator from 1 mV to 140 mV in steps of 20 mV for SI50, GI50 and GI50t and from 1 mV to 40 mV in steps of 5 mV for SI10. The variation of RMSE with amplitude is shown in Fig.~\ref{Fig:rmse_amplitude}. It is evident from Fig.~\ref{Fig:rmse_amplitude} that the RMSE decreases i.e., modal noise reduces with applied amplitude up to 20 mV for SI10 and 60 mV for all the other fibres. The modal noise and RMSE increase if the amplitude exceeds these values. Moreover, in the process of minimising RMSEs, we also lost photons due to the motion of the two mirrors. These motions cause an increase in focal spot size of the galvanometer output beam. The enlargement of spot size depends on the voltage applied. From our measurement (see Section~\ref{sec:Description_of_Galvanometer}), we found that the spot size would be about 84 $\mu$m when 60 mV voltage was applied to each of the mirrors. As this is larger than the core diameter, the overall throughput of the instrument reduces. From the colour bar of Fig.~\ref{Fig:rmse_amplitude}, the amount of loss at the RMSE minima is found to be about 30$\%$ of the peak counts for SI50 and GI50, 50$\%$ for SI10 and 5$\%$ for GI50t. Thus, we can improve the RMSEs but with a significant light-loss penalty in all fibres except GI50t. Therefore, there is a trade-off between reducing the RMSE and losing the light for most fibres tested. To further our investigation, we obtained data at two amplitudes (15 mV and 25 mV for SI10, and 20 mV and 60 mV for other fibres) as discussed in the following section.

\begin{figure*}
	\centering
	\includegraphics[width=3.4in, height=2.75in]{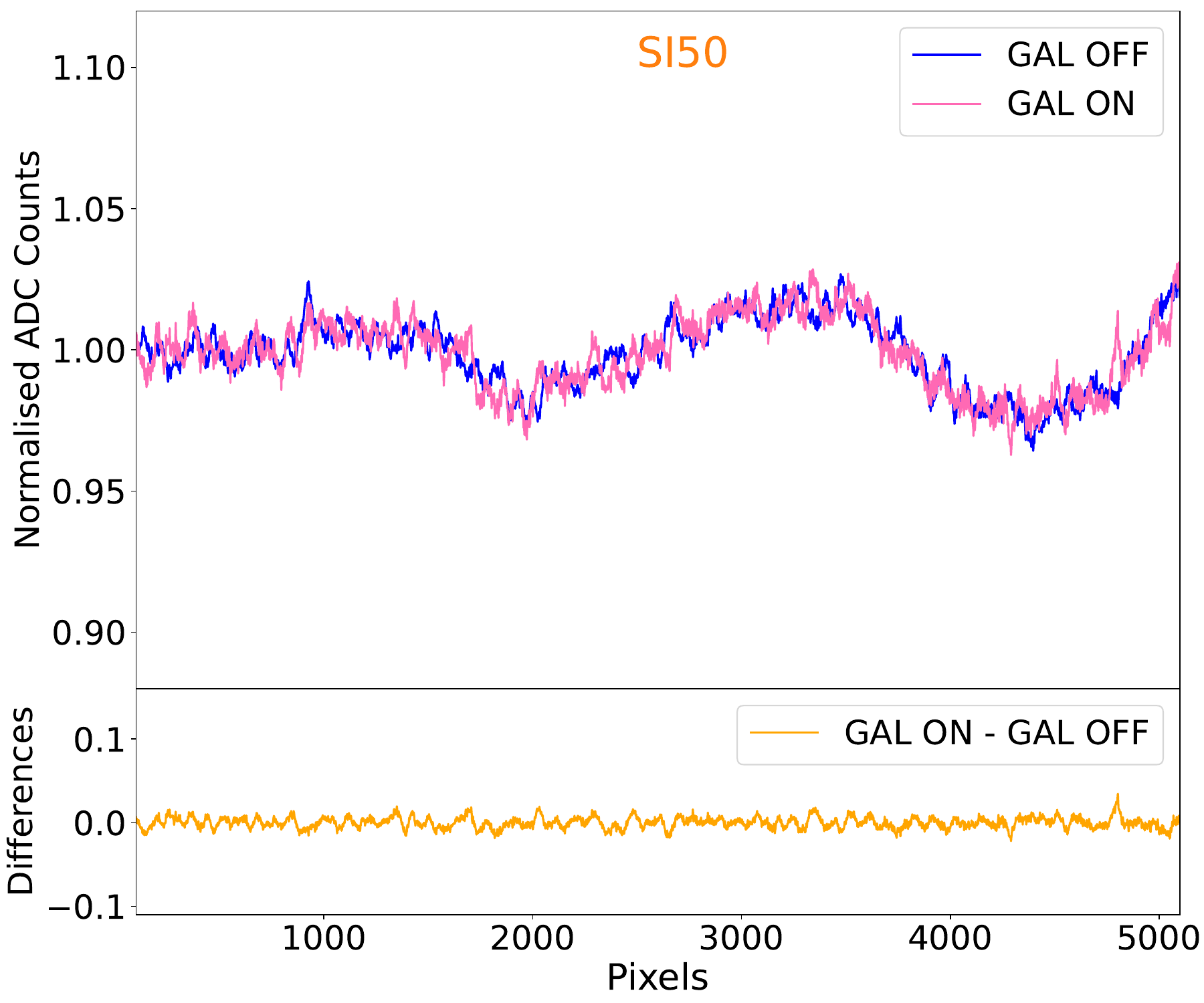}
	\includegraphics[width=3.4in, height=2.75in]{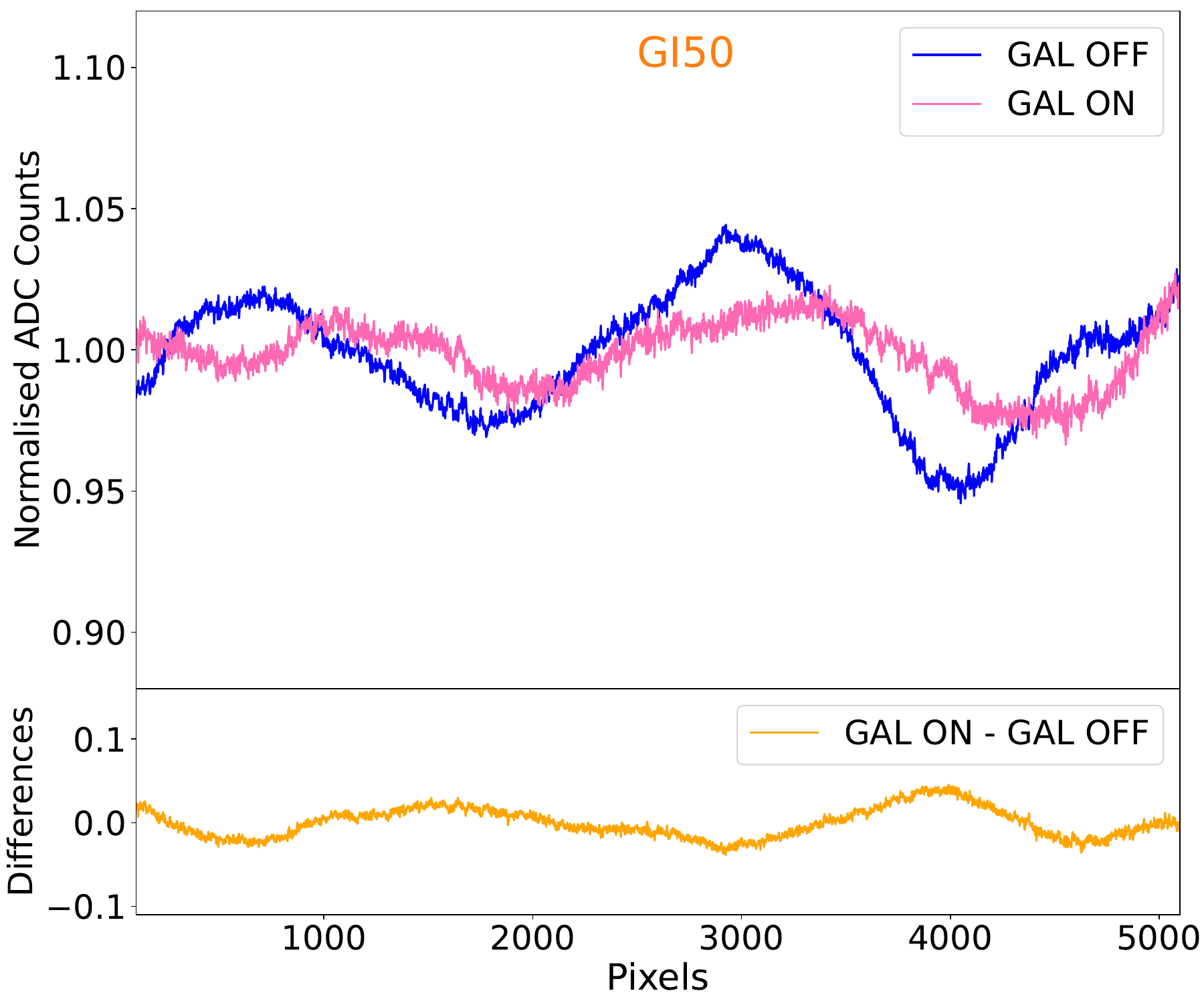}
	\includegraphics[width=3.4in, height=2.75in]{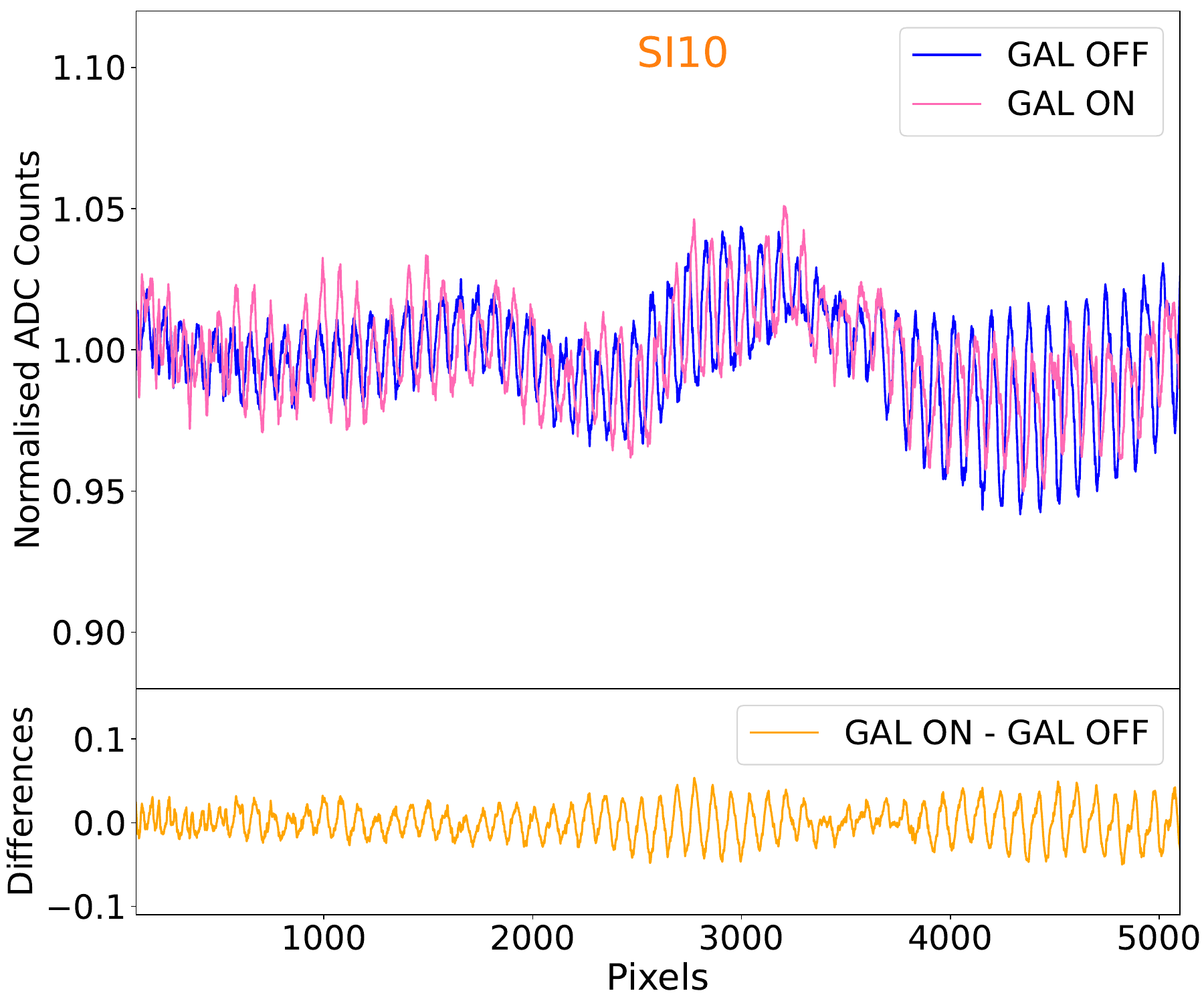}
	\includegraphics[width=3.4in, height=2.75in]{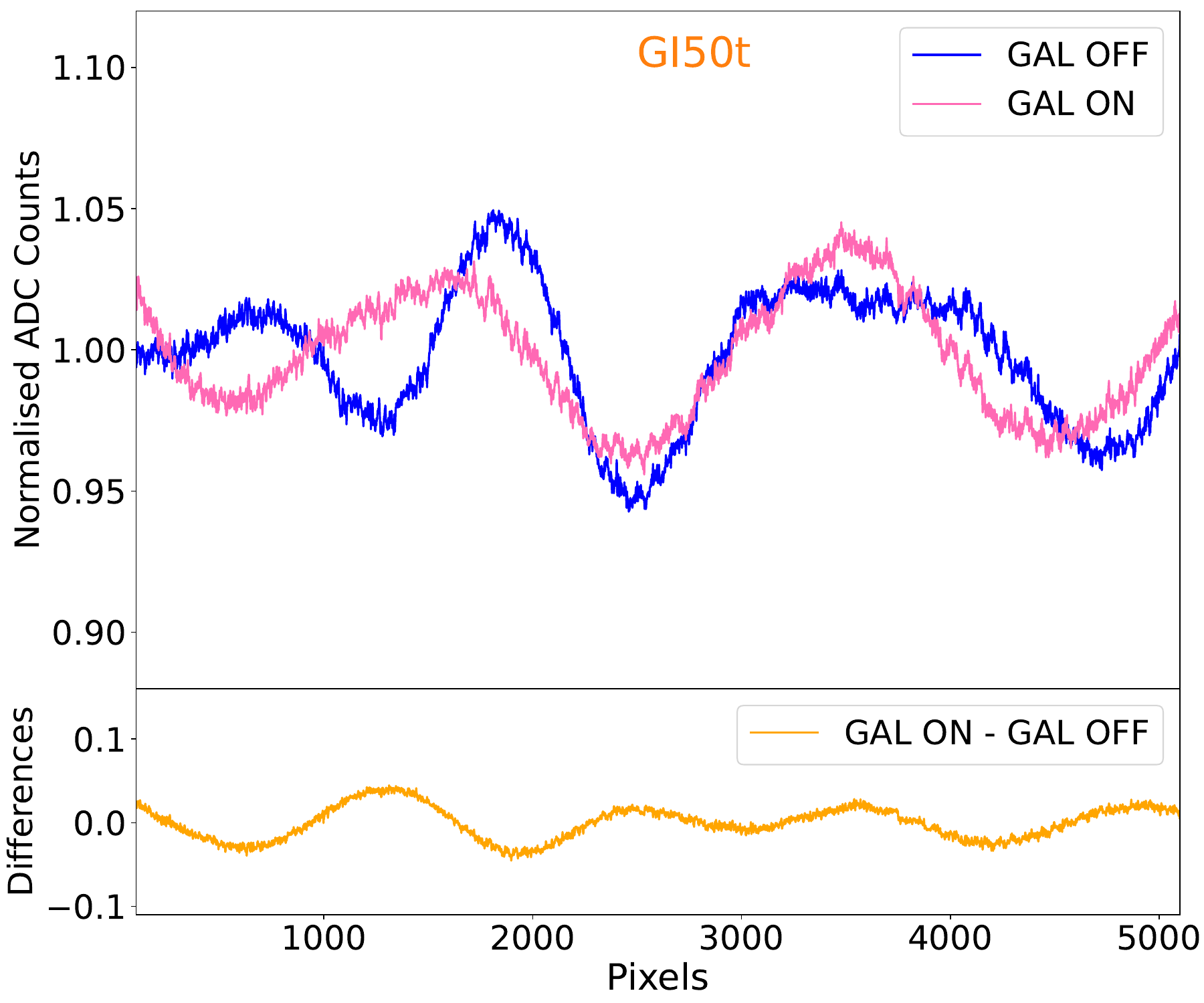}
	\caption{Extracted white light spectra of an order in GAL OFF and GAL ON conditions for different fibres (SI50, GI50, SI10, and GI50t, respectively. Applied frequencies to the voltage generator were 89 Hz and 97 Hz to CH 1 and CH 2, respectively. The applied amplitude to the voltage generator was 15 mV for SI10 and 60 mV for other fibres. The spectral regime from 100 pixels to 5000 pixels was considered for our analysis. Comparing spectrum in GAL OFF and GAL ON conditions, we can see that the spectrum are flatter when the galvanometer is on. The flat spectrum indicates the suppression of the modal noise.}
	\label{Fig:spectra_with_galvanometer}
\end{figure*}

\subsection{Investigating the efficiency of the Galvanometer in mitigating modal noise}
Once the optimum frequency and the operating amplitude of the galvanometer were set as previously described, we tested the effectiveness of our technique for all four fibres. We carried out our investigation ten times and between each observation the state of the fibre (Fibre 2) was altered. 
In order to alter the state of the fibre and consequently the mode pattern, we altered the fibre position and also manipulated the fibre by stretching, squeezing and twisting it. By doing this, we aimed to replicate all potential states of the fibre that might occur during actual sky observation in our laboratory setting. After changing the fibre state, we waited for 5 mins before starting the exposure. Our results are displayed in Fig.~\ref{Fig:rmse_obserN}. For SI50 and GI50, there is no notable change in RMSEs between the `GAL OFF' and `GAL ON' conditions. This could be due to several potential factors. For instance, SI50 and GI50 have a relatively high number of modes. Thus at any state of the fibre, the pattern of the modes is averaged out and we see less effect of modal noise on the spectra. In addition, we adopted the RMSE measurement to quantify the modal noise and our method seems to be insensitive to differentiating the modal noise between the `GAL OFF' and `GAL ON' conditions for these fibres. For GI50t, the RMSEs reduce significantly in the `GAL ON' condition in comparison to the `GAL OFF' condition. The modal noise is found to be suppressed by around 60$\%$ on average in the case of 60 mV amplitude of the applied voltage. In addition, we derived SNRs of the white light spectrum for 10 different cases as described above for GI50t in `GAL OFF' and `GAL ON' (with 60 mV) conditions following \citet{2008ASPC..394..505S}. We found an improvement of SNR of about 4$\%$ on average for `GAL ON'. An improvement of SNR implies less modal noise. For SI10, RMSEs are randomly distributed for the `GAL OFF' and `GAL ON' conditions for different observations. This could be due to the fact that the fibre is not relaxed during acquiring the data. More on fibre relaxation is discussed later in this paper (Section~\ref{sec:Fibre relaxation}). The extracted spectra in `GAL OFF' and `GAL ON' conditions for various fibres are shown in Fig.~\ref{Fig:spectra_with_galvanometer}. Comparing the spectra, significant mitigation of modal noise as well as a more symmetric modal noise pattern is seen. 
\begin{figure}
	\centering
	\includegraphics[width=3.0in, height=2.0in]{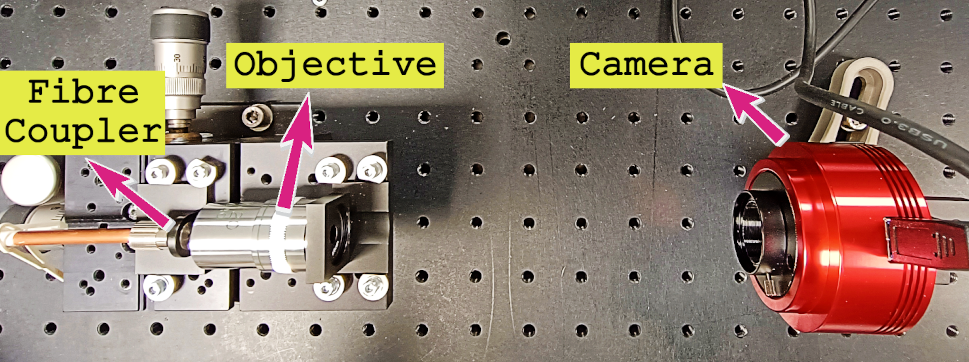}
	\caption{Illustration of microscope system setup for imaging the spatial distribution of fibres modes.}
	\label{Fig:FibreImageSetup}
\end{figure}

\begin{figure*}
	\centering
	\includegraphics[width=1.73in, height=1.5in]{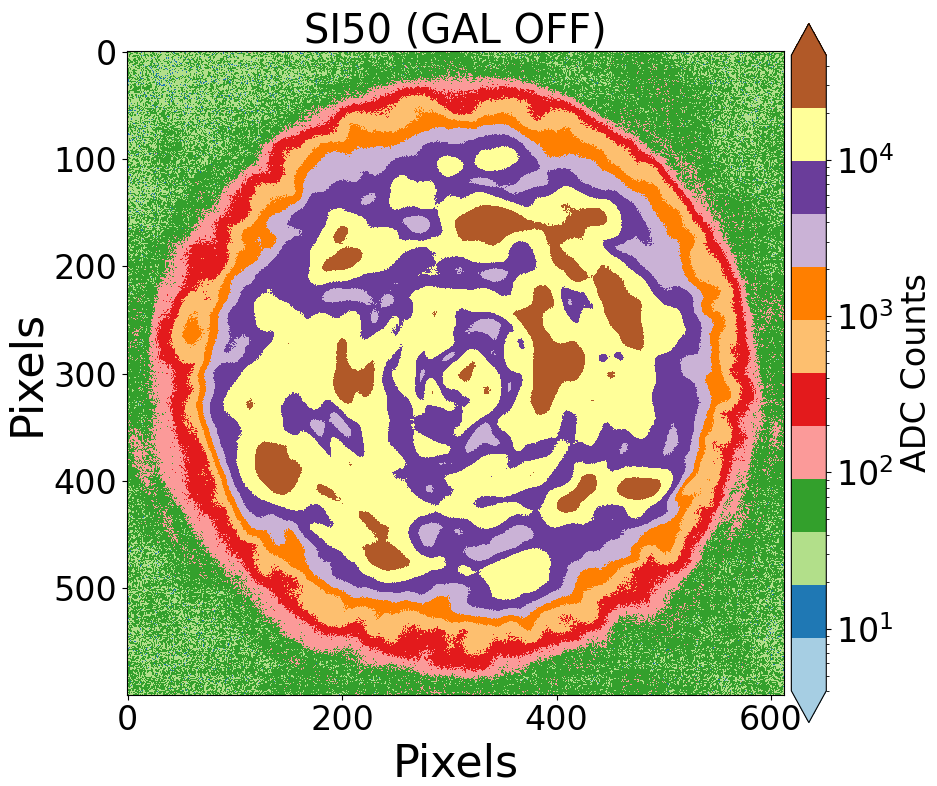}
 	\includegraphics[width=1.73in, height=1.5in]{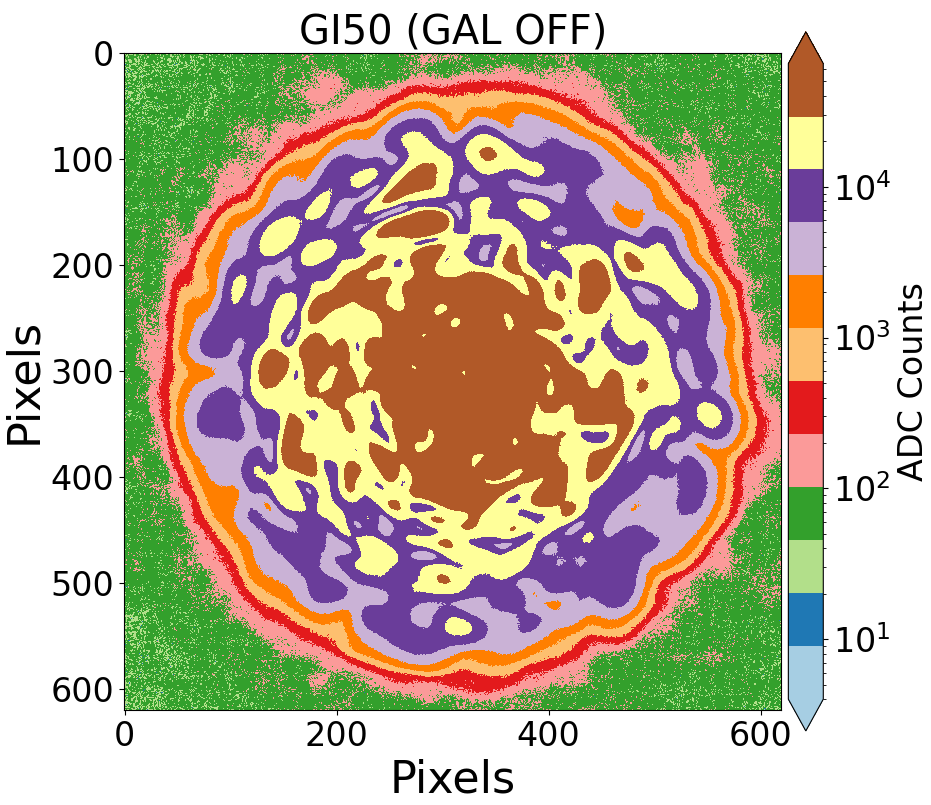}
   	\includegraphics[width=1.73in, height=1.5in]{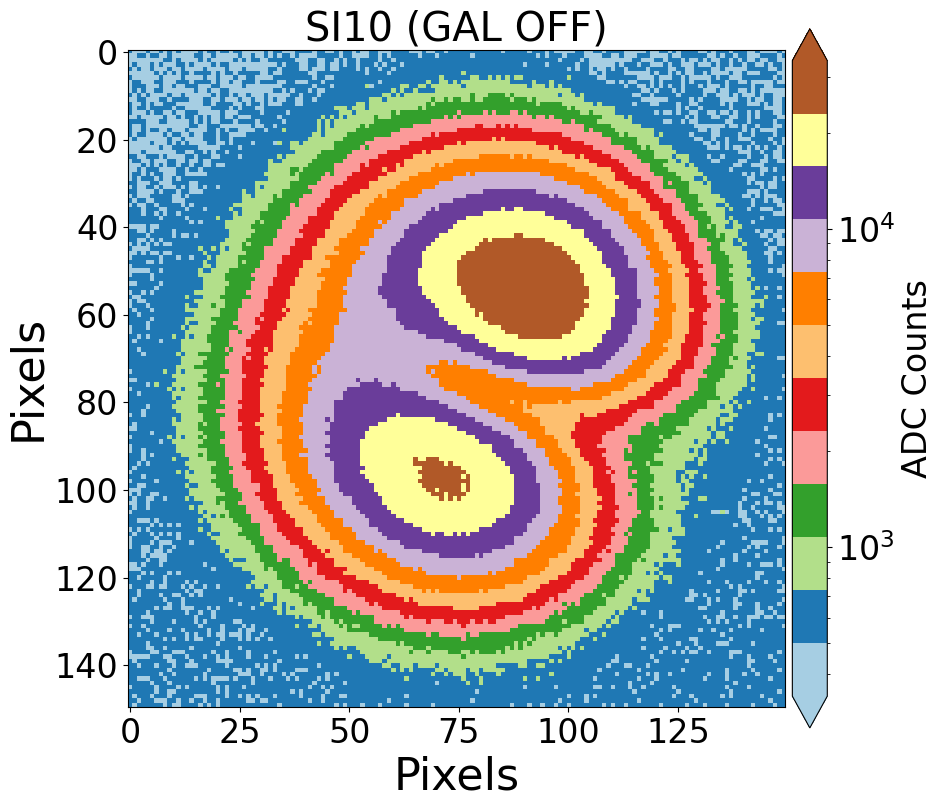}
    \includegraphics[width=1.73in, height=1.5in]{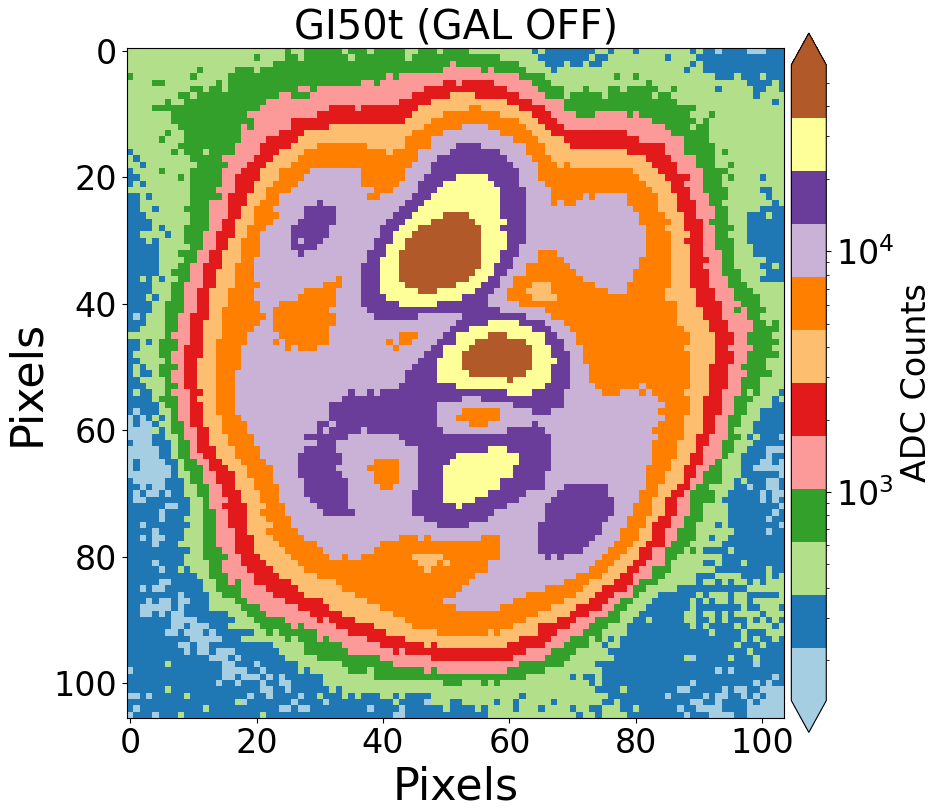}
	\includegraphics[width=1.73in, height=1.5in]{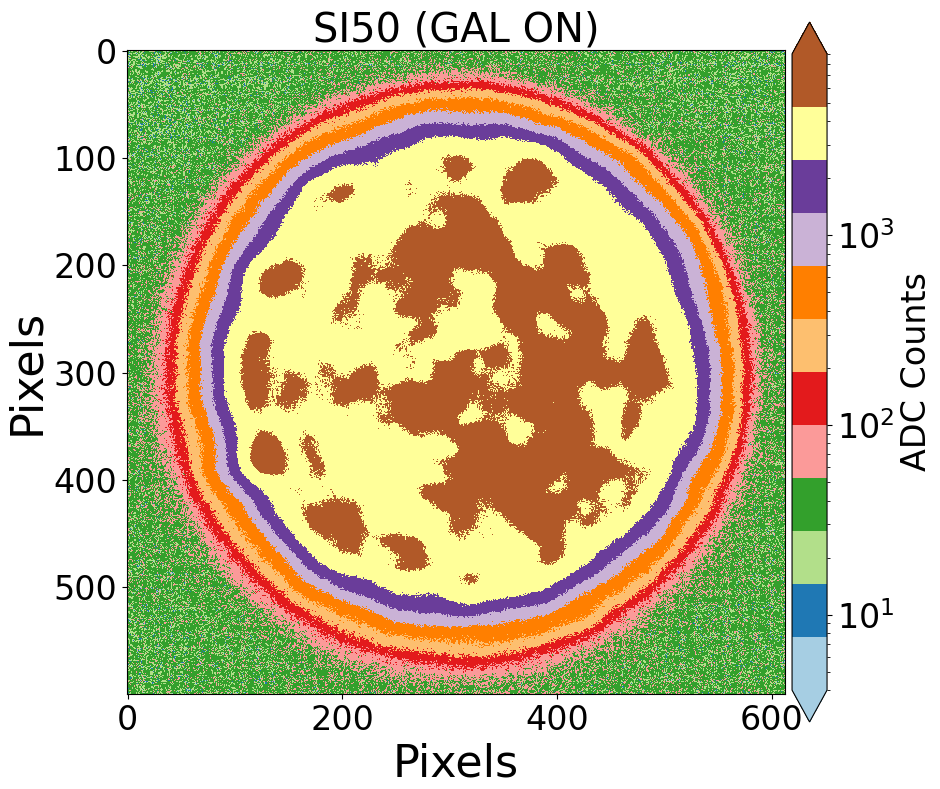}
	\includegraphics[width=1.73in, height=1.5in]{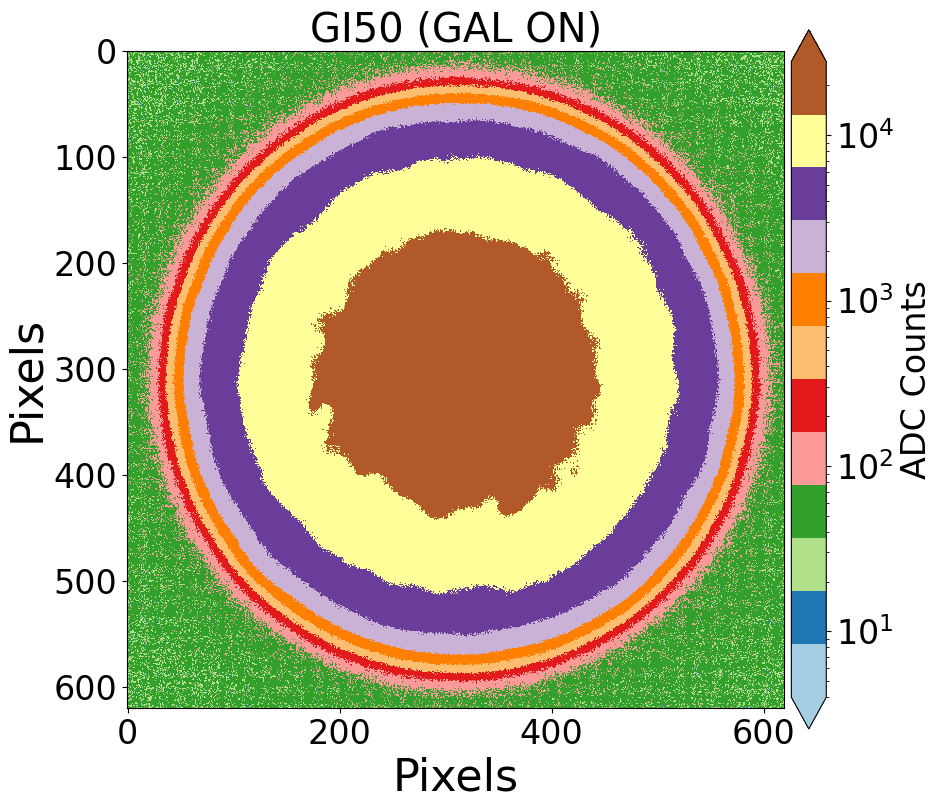}
	\includegraphics[width=1.73in, height=1.5in]{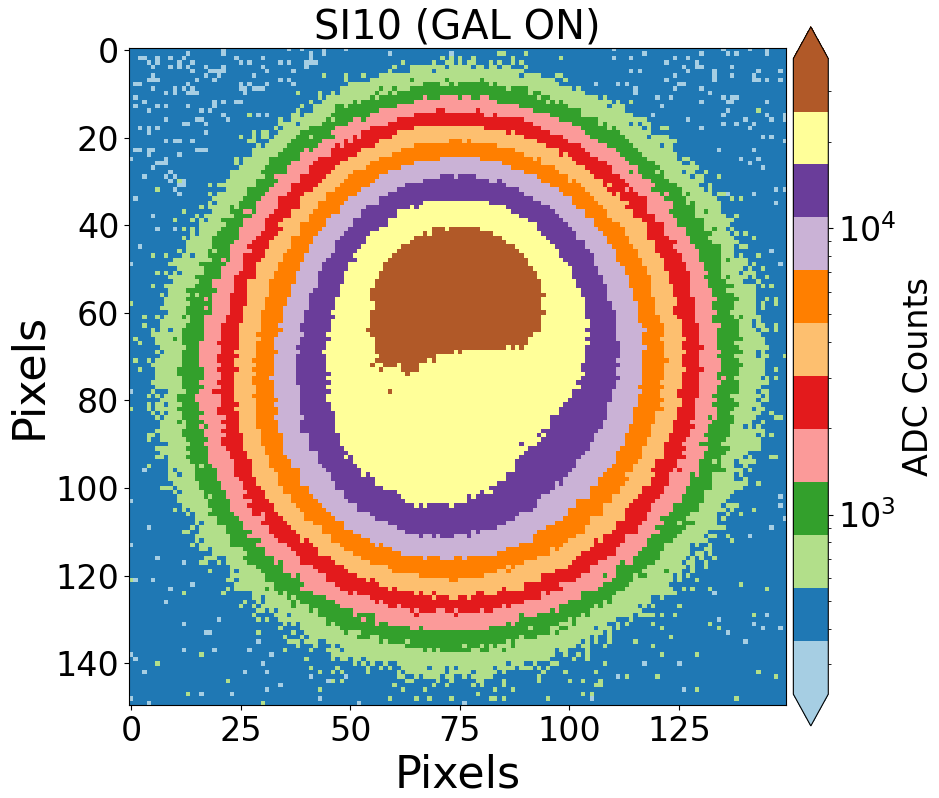} 	
	\includegraphics[width=1.73in, height=1.5in]{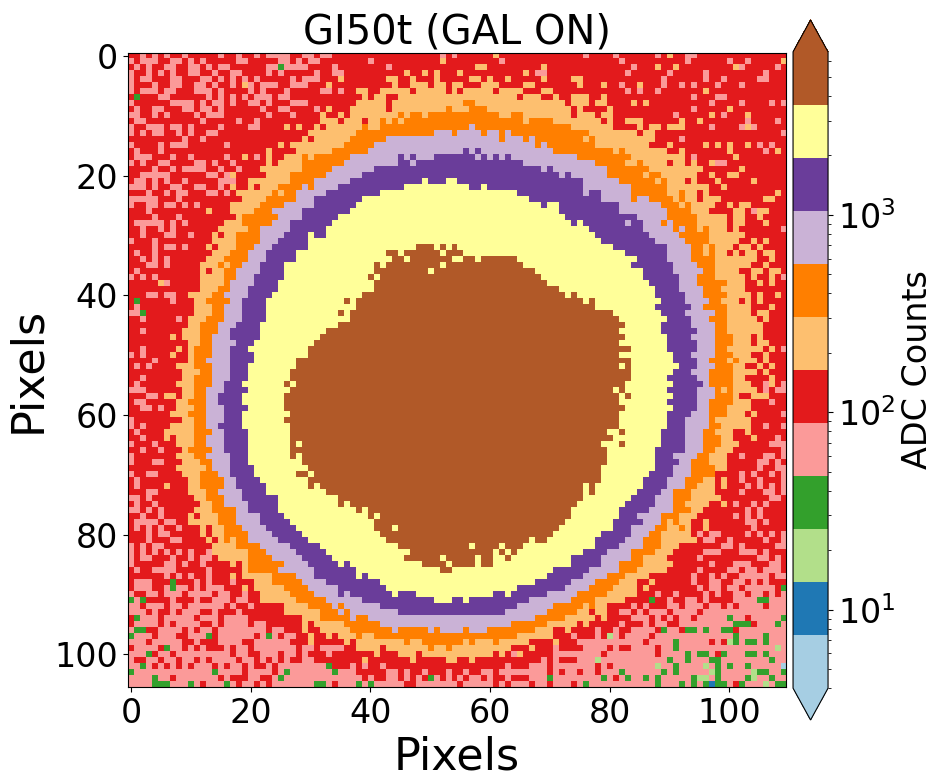}
	\caption{Comparison of images of the spatial distribution of light in various fibres in `GAL OFF' (top) and `GAL ON' (bottom). Applied frequencies to the voltage generator were 89 Hz and 97 Hz to CH 1 and CH 2, respectively. The applied amplitude to the voltage generator was 60 mV (15 mV for SI10). It is evident that the mode pattern is averaged out greatly in the `GAL ON' condition.}
	\label{Fig:fibre_image_result}
\end{figure*}
\subsection{Fibre Images with `GAL OFF' and `GAL ON'}
To demonstrate the effectiveness of the galvanometer further, we acquired images of the spatial distribution of light through our adopted fibres in `GAL OFF' and `GAL ON' conditions. For our experiment, we made a setup for the microscope optical system, where a 20x objective and a ZWO ASI camera were implemented as laid out in Fig.~\ref{Fig:FibreImageSetup}. A 635 nm red laser was used to illuminate the fibre. The complete experimental setup is similar to Fig.~\ref{Fig:Experimental_Setup} except the spectrograph (rectangular box) is replaced by the microscope. The galvanometer is operated at the optimum voltage and amplitude for various fibres at the `GAL ON' condition as mentioned in Secs~\ref{sec:frequency_measurement} and \ref{sec:amplitude_measurement}. The exposure times were 0.5 s for SI10 and 0.04 s for all other fibres. The images of the near-field speckle pattern for different fibres are displayed in Fig.~\ref{Fig:fibre_image_result}. Fig.~\ref{Fig:fibre_image_result} clearly shows that the addition of the galvanometer significantly improves the speckle uniformity. We note, however, a relatively high background for GI50t. This effect could possibly be due to the fact that cladding mode suppression for other fibres was much better than the tapered fibre used for this test.

\subsection{Fibre relaxation} \label{sec:Fibre relaxation}
\begin{figure}
	\centering
	\includegraphics[width=3.2in, height=3.0in]{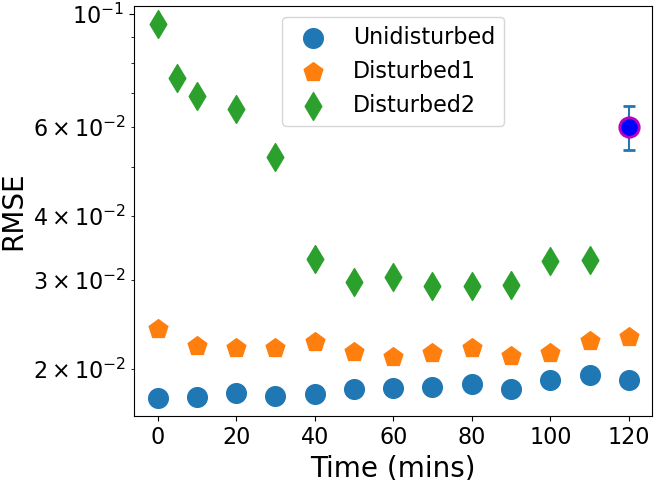}
	\caption{Variation of RMSE with Time to test the fibre relaxation for SI10 with `GAL OFF condition'. The RMSE variation for three states of the fibre is presented. In the first case (blue symbols), the fibre was connected from the white light source to the spectrograph and allowed to rest overnight. This state of the fibre is labelled as `Undisturbed'. In the second case (orange symbols, `Disturbed1'), the position of the fibre was altered by a small amount i.e., the fibre was disconnected from the fibre coupler of the white light source and reconnected. In the third case (green symbols), the fibre was stretched along its length and allowed to relax at the same position. This state of the fibre was termed `Disturbed2'. It is evident that the RMSE value decreases with time representing the stabilization of the mode pattern. Blue error bar at the top represents typical RMSE measurement error.}
	\label{Fig:fibre_relaxation}
\end{figure}
We extended our investigation further to explore the effect of fibre relaxation for SI10. The variation of RMSE with time for three different cases is shown in Fig.~\ref{Fig:fibre_relaxation}. For the first case (`Undisturbed', blue circles), the fibre was connected between the white light source and the spectrograph and allowed to rest overnight. No notable change in RMSE values with time was evident in this case. As the fibre was rested overnight, it can be considered to be in a relaxed condition, and thus, there is no distinct change in mode distribution between exposures. In the second case (`Disturbed1', orange symbols), the fibre was disconnected from and reconnected to the fibre coupler of the white light source. Change in RMSE values were apparent in comparison to the former case. However, the variation of RMSE with time was insignificant. The reason for the change is probably due to the redistribution of modes because of the change in fibre coupling and external disturbance. In the third case (Disturbed2, green markers), the fibre was stretched along its length and then allowed to relax in the same condition. In this case, the RMSE was found to change significantly with time for the first 50 mins. After 50 mins, the change is insignificant. As the experiment was carried out in a general purpose temperature controlled laboratory environment, large temperature variations can be ruled out as a major contributor. 
Therefore, we believe that small stresses and strains due to externally applied forces are the reason for these changes. When the fibre was stretched, the applied external force disrupts the mechanical equilibrium in the fibre and as a result, the RMSE value increases. When the fibre was allowed to relax, it tends to back to an unstressed and unstrained state at room temperature and so RMSE decreases. However, the final relaxed state and hence the propagating mode patterns depends on the incident electric field, wavelength, fibre stresses and strains \citep{Minardi2021}.
The observed variation of RMSE over time in the case of Disturbed2 might not be as prominent in fibres with a large core. Due to the large number of modes, the mode pattern tends to average out quickly. It is important to note that for small core fibre such as 10 $\mu$m fibre, it is crucial to exercise additional caution against the movement of the fibre during the experiment. Additional experiments are required to fully understand this effect. Recent studies showed that thermal and mechanical stabilization of a multimode fibre can greatly improve its' long-term stability (\citealt{Redding:14, Liew:16, Cao:23}). Furthermore, the most satisfactory explanation of our galvanometer-based modal noise mitigation method and fibre relaxation might be obtained using a complex wave-optics simulation model such as COMSOL\footnote{\url{https://uk.comsol.com/}}. For such calculations, we need to treat each mode of the fibre independently with random variations of phases and mode patterns. Therefore, the model calculation will be computationally expensive with significant computer time. Thus, the detailed theoretical study is beyond the scope of this paper.

\section{Summary}
In this paper, we have demonstrated the use of commercial off-the-shelf components to mitigate modal noise for fibre-fed high-resolution spectrographs. Our method uses a galvanometer designed to randomize the mode distribution while the light passes through it. The galvanometer uses two motorized rotating flat mirrors to mix the modes. The mode mixing can be controlled by setting up the frequency and amplitude of the applied sinusoidal signal through a voltage generator connected to each of the mirrors. To carry out our experiment, we used four different fibres: 50 $\mu$m step-index, 50 $\mu$m graded-index, 10 $\mu$m step-index, and a combination of 50 $\mu$m graded-index and 5:1 tapered fibre. We obtained white light spectra using a high-resolution spectrograph for our experiment and used RMSE measurement as a proxy for modal noise quantification. We first investigated the optimal frequency and amplitude of signal applied for efficient operation of the galvanometer. No significant frequency dependence on RMSE values was seen. However, the amplitude of the signal applied to the galvanometer did affect the RMSE and we used our measurement to find the optimum amplitude and maximise the galvanometer's efficiency. Once the frequency and amplitude conditions were found, we repeated our experiment multiple times both with and without agitation by the galvanometer and showed that our technique was capable of mitigating modal noise efficiently. Among all the tested fibres, the graded-index fibre combined with the tapered fibre (GI50t) was found to mitigate modal noise on an average by 60$\%$ with a output light-loss penalty of only 5$\%$. In addition, we obtained images of the mode distribution for all fibres both with and without galvanometer agitation. We found that the use of a galvanometer significantly improves the uniformity of light illumination at the fibre output. 
Our technique offers benefits over other methods because of its simplicity, cost-effectiveness, and ease of implementation. We do not need to move the fibre during the observation. Thus, our methods is free from any mechanical agitation. It also does not compromise the lifetime of the fibre which might be the case if intense mechanical agitation methods are employed. However, there is a trade-off between improvement in modal noise and loss of light which limits the performance of the galvanometer and the throughput efficiency of the spectrograph. Additionally, we found that stressing the fibre with an external force can destabilise the spatial distribution of modes especially in small core fibres such as 10 $\mu$m step-index fibre. It takes a significantly long time ($\sim$1 hr) for the fibre to return to a relaxed state. Therefore, such fibres need to be handled very carefully when used for spectroscopic observations.

\section*{Acknowledgements}
The authors are thankful to the reviewers for their critical and valuable comments on the original version of the manuscript, which helped us to improve the paper. We are very grateful to Tim Birks at University of Bath for his support in construction and supply of custom GI50t fibre. We are thankful for support of this research work through Newton fund grants from STFC under project identification number ST/P005667/1, ST/R006598/1 and ST/T007311/1. 

\section*{Data Availability}
The data underlying this article will be shared on reasonable request to the corresponding author.

 



\bibliographystyle{rasti}
\bibliography{ModalNoisePaper_rastiV_final}

\begin{thebibliography}{49}
\expandafter\ifx\csname natexlab\endcsname\relax\def\natexlab#1{#1}\fi

\bibitem[{Allen} et~al.(2018){Allen}, {Wright}, {Rajagopal}, {Santoro},
  {Liang}, {Timmerman}, {Christensen}, {Hunting}, {Wolf}, {Jaehnig},
  {Percival}, {Smith}, {Mahadevan}, {Hearty}, {Bender}, {Blake}, {Logsdon},
  {Akeson}, {Capps}, {Callas}, {Willems}, {McElwain}, {McElwain}, {Basten},
  {Monson}, {Stefansson}, {Ramsey}, {Ninan}, {Blakeslee}, {Kaplan},
  {Halverson}, {Roy}, {Terrien}, {Robertson}, {Schwab}, {Rud}, \&
  {Kanodia}]{2018AAS...23124608A}
{Allen}, L.~E., {Wright}, J., {Rajagopal}, J., {Santoro}, F., {Liang}, M.,
  {Timmerman}, E., {Christensen}, R., {Hunting}, E., {Wolf}, M., {Jaehnig}, K.,
  {Percival}, J., {Smith}, M., {Mahadevan}, S., {Hearty}, F., {Bender}, C.,
  {Blake}, C., {Logsdon}, S.~E., {Akeson}, R., {Capps}, R.~W., {Callas}, J.,
  {Willems}, P.~A., {McElwain}, M., {McElwain}, M.~W., {Basten}, F., {Monson},
  A., {Stefansson}, G., {Ramsey}, L., {Ninan}, J., {Blakeslee}, S., {Kaplan},
  K., {Halverson}, S., {Roy}, A., {Terrien}, R., {Robertson}, P., {Schwab}, C.,
  {Rud}, M., \& {Kanodia}, S., 2018.
\newblock {NEID: A High Precision Radial Velocity Spectrograph for the WIYN
  3.5-m Telescope}, in {\em American Astronomical Society Meeting Abstracts
  \#231\/}, vol. 231 of {\bf American Astronomical Society Meeting Abstracts},
  p. 246.08.

\bibitem[{Artigau} et~al.(2014){Artigau}, {Kouach}, {Donati}, {Doyon},
  {Delfosse}, {Baratchart}, {Lacombe}, {Moutou}, {Rabou}, {Par{\`e}s},
  {Micheau}, {Thibault}, {Reshetov}, {Dubois}, {Hernandez}, {Vall{\'e}e},
  {Wang}, {Dolon}, {Pepe}, {Bouchy}, {Striebig}, {H{\'e}nault}, {Loop},
  {Saddlemyer}, {Barrick}, {Vermeulen}, {Dupieux}, {H{\'e}brard}, {Boisse},
  {Martioli}, {Alencar}, {do Nascimento}, \& {Figueira}]{2014SPIE.9147E..15A}
{Artigau}, {\'E}., {Kouach}, D., {Donati}, J.-F., {Doyon}, R., {Delfosse}, X.,
  {Baratchart}, S., {Lacombe}, M., {Moutou}, C., {Rabou}, P., {Par{\`e}s},
  L.~P., {Micheau}, Y., {Thibault}, S., {Reshetov}, V.~A., {Dubois}, B.,
  {Hernandez}, O., {Vall{\'e}e}, P., {Wang}, S.-Y., {Dolon}, F., {Pepe}, F.~A.,
  {Bouchy}, F., {Striebig}, N., {H{\'e}nault}, F., {Loop}, D., {Saddlemyer},
  L., {Barrick}, G., {Vermeulen}, T., {Dupieux}, M., {H{\'e}brard}, G.,
  {Boisse}, I., {Martioli}, E., {Alencar}, S. H.~P., {do Nascimento}, J.-D., \&
  {Figueira}, P., 2014.
\newblock {SPIRou: the near-infrared spectropolarimeter/high-precision
  velocimeter for the Canada-France-Hawaii telescope}, in {\em Ground-based and
  Airborne Instrumentation for Astronomy V\/}, vol. 9147 of {\bf Society of
  Photo-Optical Instrumentation Engineers (SPIE) Conference Series}, p. 914715.

\bibitem[{Avila}(1988)]{1988ASPC....3...63A}
{Avila}, G., 1988.
\newblock {Tests of optical fibres for astronomical instrumentation at ESO.},
  in {\em Fiber Optics in Astronomy\/}, vol.~3 of {\bf Astronomical Society of
  the Pacific Conference Series}, pp. 63--73.

\bibitem[{Baudrand} \& {Walker}(2001)]{2001PASP..113..851B}
{Baudrand}, J. \& {Walker}, G. A.~H., 2001.
\newblock {Modal Noise in High-Resolution, Fiber-fed Spectra: A Study and
  Simple Cure}, {\it \pasp\/}, {\bf 113}(785), 851--858.

\bibitem[{Birks} et~al.(2015){Birks}, {Gris-S{\'a}nchez}, {Yerolatsitis},
  {Leon-Saval}, \& {Thomson}]{2015AdOP....7..107B}
{Birks}, T.~A., {Gris-S{\'a}nchez}, I., {Yerolatsitis}, S., {Leon-Saval},
  S.~G., \& {Thomson}, R.~R., 2015.
\newblock {The photonic lantern}, {\it Advances in Optics and Photonics\/},
  {\bf 7}(2), 107.

\bibitem[{Blake} et~al.(2015){Blake}, {Johnson}, {Plavchan}, {Sliski},
  {Wittenmyer}, {Eastman}, \& {Barnes}]{2015AAS...22525732B}
{Blake}, C., {Johnson}, J., {Plavchan}, P., {Sliski}, D., {Wittenmyer}, R.~A.,
  {Eastman}, J.~D., \& {Barnes}, S., 2015.
\newblock {MINERVA-Red: A Census of Planets Orbiting the Nearest Low-mass Stars
  to the Sun}, in {\em American Astronomical Society Meeting Abstracts
  \#225\/}, vol. 225 of {\bf American Astronomical Society Meeting Abstracts},
  p. 257.32.

\bibitem[{Bouchy} et~al.(2017){Bouchy}, {Doyon}, {Artigau}, {Melo},
  {Hernandez}, {Wildi}, {Delfosse}, {Lovis}, {Figueira}, {Canto Martins},
  {Gonz{\'a}lez Hern{\'a}ndez}, {Thibault}, {Reshetov}, {Pepe}, {Santos}, {de
  Medeiros}, {Rebolo}, {Abreu}, {Adibekyan}, {Bandy}, {Benz}, {Blind},
  {Bohlender}, {Boisse}, {Bovay}, {Broeg}, {Brousseau}, {Cabral}, {Chazelas},
  {Cloutier}, {Coelho}, {Conod}, {Cumming}, {Delabre}, {Genolet}, {Hagelberg},
  {Jayawardhana}, {K{\"a}ufl}, {Lafreni{\`e}re}, {de Castro Le{\~a}o}, {Malo},
  {de Medeiros Martins}, {Matthews}, {Metchev}, {Oshagh}, {Ouellet}, {Parro},
  {Rasilla Pi{\~n}eiro}, {Santos}, {Sarajlic}, {Segovia}, {Sordet}, {Udry},
  {Valencia}, {Vall{\'e}e}, {Venn}, {Wade}, \&
  {Saddlemyer}]{2017Msngr.169...21B}
{Bouchy}, F., {Doyon}, R., {Artigau}, {\'E}., {Melo}, C., {Hernandez}, O.,
  {Wildi}, F., {Delfosse}, X., {Lovis}, C., {Figueira}, P., {Canto Martins},
  B.~L.~., {Gonz{\'a}lez Hern{\'a}ndez}, J.~I.~., {Thibault}, S., {Reshetov},
  V., {Pepe}, F., {Santos}, N.~C., {de Medeiros}, J.~R.~., {Rebolo}, R.,
  {Abreu}, M., {Adibekyan}, V.~Z., {Bandy}, T., {Benz}, W., {Blind}, N.,
  {Bohlender}, D., {Boisse}, I., {Bovay}, S., {Broeg}, C., {Brousseau}, D.,
  {Cabral}, A., {Chazelas}, B., {Cloutier}, R., {Coelho}, J., {Conod}, U.,
  {Cumming}, A., {Delabre}, B., {Genolet}, L., {Hagelberg}, J., {Jayawardhana},
  R., {K{\"a}ufl}, H.~U., {Lafreni{\`e}re}, D., {de Castro Le{\~a}o}, I.~.,
  {Malo}, L., {de Medeiros Martins}, A.~., {Matthews}, J.~M., {Metchev}, S.,
  {Oshagh}, M., {Ouellet}, M., {Parro}, V.~C., {Rasilla Pi{\~n}eiro}, J.~L.~.,
  {Santos}, P., {Sarajlic}, M., {Segovia}, A., {Sordet}, M., {Udry}, S.,
  {Valencia}, D., {Vall{\'e}e}, P., {Venn}, K., {Wade}, G.~A., \& {Saddlemyer},
  L., 2017.
\newblock {Near-InfraRed Planet Searcher to Join HARPS on the ESO 3.6-metre
  Telescope}, {\it The Messenger\/}, {\bf 169}, 21--27.

\bibitem[Cao et~al.(2023)Cao, \v{C}i\v{z}m\'{a}r, Turtaev, Tyc, \&
  Rotter]{Cao:23}
Cao, H., \v{C}i\v{z}m\'{a}r, T., Turtaev, S., Tyc, T., \& Rotter, S., 2023.
\newblock Controlling light propagation in multimode fibers for imaging,
  spectroscopy, and beyond, {\it Adv. Opt. Photon.\/}, {\bf 15}(2), 524--612.

\bibitem[{Chakraborty} et~al.(2014){Chakraborty}, {Mahadevan}, {Roy}, {Dixit},
  {Richardson}, {Dongre}, {Pathan}, {Chaturvedi}, {Shah}, {Ubale}, \&
  {Anandarao}]{2014PASP..126..133C}
{Chakraborty}, A., {Mahadevan}, S., {Roy}, A., {Dixit}, V., {Richardson},
  E.~H., {Dongre}, V., {Pathan}, F.~M., {Chaturvedi}, P., {Shah}, V., {Ubale},
  G.~P., \& {Anandarao}, B.~G., 2014.
\newblock {The PRL Stabilized High-Resolution Echelle Fiber-fed Spectrograph:
  Instrument Description and First Radial Velocity Results}, {\it \pasp\/},
  {\bf 126}(936), 133.

\bibitem[Choochalerm et~al.(2021)Choochalerm, Martin, Jones, Errmann,
  Yerolatsitis, Wright, \& Buisset]{CHOOCHALERM2021102632}
Choochalerm, P., Martin, W.~E., Jones, H.~R., Errmann, R., Yerolatsitis, S.,
  Wright, T.~A., \& Buisset, C., 2021.
\newblock Transmission properties of tapered optical fibres: Simulations and
  experimental measurements, {\it Optical Fiber Technology\/}, {\bf 66},
  102632.

\bibitem[Choochalerm et~al.(2023)Choochalerm, Martin, Jones, Usher, Wright, \&
  Yerolatsitis]{CHOOCHALERM2023103140}
Choochalerm, P., Martin, W.~E., Jones, H.~R., Usher, S., Wright, T.~A., \&
  Yerolatsitis, S., 2023.
\newblock Incoherent light in tapered graded-index fibre: A study of
  transmission and modal noise, {\it Optical Fiber Technology\/}, {\bf 75},
  103140.

\bibitem[{Crepp}(2014)]{2014Sci...346..809C}
{Crepp}, J.~R., 2014.
\newblock {Improving planet-finding spectrometers}, {\it Science\/}, {\bf
  346}(6211), 809--810.

\bibitem[{Crepp} et~al.(2016){Crepp}, {Crass}, {King}, {Bechter}, {Bechter},
  {Ketterer}, {Reynolds}, {Hinz}, {Kopon}, {Cavalieri}, {Fantano}, {Koca},
  {Onuma}, {Stapelfeldt}, {Thomes}, {Wall}, {Macenka}, {McGuire}, {Korniski},
  {Zugby}, {Eisner}, {Gaudi}, {Hearty}, {Kratter}, {Kuchner}, {Micela},
  {Nelson}, {Pagano}, {Quirrenbach}, {Schwab}, {Skrutskie}, {Sozzetti},
  {Woodward}, \& {Zhao}]{2016SPIE.9908E..19C}
{Crepp}, J.~R., {Crass}, J., {King}, D., {Bechter}, A., {Bechter}, E.,
  {Ketterer}, R., {Reynolds}, R., {Hinz}, P., {Kopon}, D., {Cavalieri}, D.,
  {Fantano}, L., {Koca}, C., {Onuma}, E., {Stapelfeldt}, K., {Thomes}, J.,
  {Wall}, S., {Macenka}, S., {McGuire}, J., {Korniski}, R., {Zugby}, L.,
  {Eisner}, J., {Gaudi}, B.~S., {Hearty}, F., {Kratter}, K., {Kuchner}, M.,
  {Micela}, G., {Nelson}, M., {Pagano}, I., {Quirrenbach}, A., {Schwab}, C.,
  {Skrutskie}, M., {Sozzetti}, A., {Woodward}, C., \& {Zhao}, B., 2016.
\newblock {iLocater: a diffraction-limited Doppler spectrometer for the Large
  Binocular Telescope}, in {\em Ground-based and Airborne Instrumentation for
  Astronomy VI\/}, vol. 9908 of {\bf Society of Photo-Optical Instrumentation
  Engineers (SPIE) Conference Series}, p. 990819.

\bibitem[Epworth(1978)]{Epworth:1978}
Epworth, R., 1978.
\newblock The phenomenon of modal noise in analogue and digital optical fibre
  systems.

\bibitem[{Fischer} et~al.(2016){Fischer}, {Anglada-Escude}, {Arriagada},
  {Baluev}, {Bean}, {Bouchy}, {Buchhave}, {Carroll}, {Chakraborty}, {Crepp},
  {Dawson}, {Diddams}, {Dumusque}, {Eastman}, {Endl}, {Figueira}, {Ford},
  {Foreman-Mackey}, {Fournier}, {F{\H{u}}r{\'e}sz}, {Gaudi}, {Gregory},
  {Grundahl}, {Hatzes}, {H{\'e}brard}, {Herrero}, {Hogg}, {Howard}, {Johnson},
  {Jorden}, {Jurgenson}, {Latham}, {Laughlin}, {Loredo}, {Lovis}, {Mahadevan},
  {McCracken}, {Pepe}, {Perez}, {Phillips}, {Plavchan}, {Prato}, {Quirrenbach},
  {Reiners}, {Robertson}, {Santos}, {Sawyer}, {Segransan}, {Sozzetti},
  {Steinmetz}, {Szentgyorgyi}, {Udry}, {Valenti}, {Wang}, {Wittenmyer}, \&
  {Wright}]{2016PASP..128f6001F}
{Fischer}, D.~A., {Anglada-Escude}, G., {Arriagada}, P., {Baluev}, R.~V.,
  {Bean}, J.~L., {Bouchy}, F., {Buchhave}, L.~A., {Carroll}, T., {Chakraborty},
  A., {Crepp}, J.~R., {Dawson}, R.~I., {Diddams}, S.~A., {Dumusque}, X.,
  {Eastman}, J.~D., {Endl}, M., {Figueira}, P., {Ford}, E.~B.,
  {Foreman-Mackey}, D., {Fournier}, P., {F{\H{u}}r{\'e}sz}, G., {Gaudi}, B.~S.,
  {Gregory}, P.~C., {Grundahl}, F., {Hatzes}, A.~P., {H{\'e}brard}, G.,
  {Herrero}, E., {Hogg}, D.~W., {Howard}, A.~W., {Johnson}, J.~A., {Jorden},
  P., {Jurgenson}, C.~A., {Latham}, D.~W., {Laughlin}, G., {Loredo}, T.~J.,
  {Lovis}, C., {Mahadevan}, S., {McCracken}, T.~M., {Pepe}, F., {Perez}, M.,
  {Phillips}, D.~F., {Plavchan}, P.~P., {Prato}, L., {Quirrenbach}, A.,
  {Reiners}, A., {Robertson}, P., {Santos}, N.~C., {Sawyer}, D., {Segransan},
  D., {Sozzetti}, A., {Steinmetz}, T., {Szentgyorgyi}, A., {Udry}, S.,
  {Valenti}, J.~A., {Wang}, S.~X., {Wittenmyer}, R.~A., \& {Wright}, J.~T.,
  2016.
\newblock {State of the Field: Extreme Precision Radial Velocities}, {\it
  \pasp\/}, {\bf 128}(964), 066001.

\bibitem[Floris et~al.(2020)Floris, {de Hon}, Bolhaar, \&
  Smink]{FLORIS2020102116}
Floris, S.~J., {de Hon}, B.~P., Bolhaar, T., \& Smink, R.~W., 2020.
\newblock A geometrical optics approach to encircled-flux compliant source
  modeling for multi-mode fiber illumination and connection attenuation, {\it
  Optical Fiber Technology\/}, {\bf 54}, 102116.

\bibitem[{Halverson} et~al.(2015{\natexlab{a}}){Halverson}, {Roy}, {Mahadevan},
  {Ramsey}, {Levi}, {Schwab}, {Hearty}, \& {MacDonald}]{2015ApJ...806...61H}
{Halverson}, S., {Roy}, A., {Mahadevan}, S., {Ramsey}, L., {Levi}, E.,
  {Schwab}, C., {Hearty}, F., \& {MacDonald}, N., 2015{\natexlab{a}}.
\newblock {An Efficient, Compact, and Versatile Fiber Double Scrambler for High
  Precision Radial Velocity Instruments}, {\it \apj\/}, {\bf 806}(1), 61.

\bibitem[{Halverson} et~al.(2015{\natexlab{b}}){Halverson}, {Roy}, {Mahadevan},
  \& {Schwab}]{2015ApJ...814L..22H}
{Halverson}, S., {Roy}, A., {Mahadevan}, S., \& {Schwab}, C.,
  2015{\natexlab{b}}.
\newblock {``Modal Noise'' in Single-mode Fibers: A Cautionary Note for High
  Precision Radial Velocity Instruments}, {\it \apjl\/}, {\bf 814}(2), L22.

\bibitem[Hill et~al.(1980)Hill, Tremblay, \& Kawasaki]{Hill:80}
Hill, K.~O., Tremblay, Y., \& Kawasaki, B.~S., 1980.
\newblock Modal noise in multimode fiber links: theory and experiment, {\it
  Opt. Lett.\/}, {\bf 5}(6), 270--272.

\bibitem[{Hunter} \& {Ramsey}(1992)]{1992PASP..104.1244H}
{Hunter}, T.~R. \& {Ramsey}, L.~W., 1992.
\newblock {Scrambling Properties of Optical Fibers and the Performance of a
  Double Scrambler}, {\it \pasp\/}, {\bf 104}, 1244.

\bibitem[{Ishizuka} et~al.(2018){Ishizuka}, {Kotani}, {Nishikawa}, {Kurokawa},
  {Mori}, {Kokubo}, \& {Tamura}]{2018PASP..130f5003I}
{Ishizuka}, M., {Kotani}, T., {Nishikawa}, J., {Kurokawa}, T., {Mori}, T.,
  {Kokubo}, T., \& {Tamura}, M., 2018.
\newblock {Fiber Mode Scrambler for the Subaru Infrared Doppler Instrument
  (IRD)}, {\it \pasp\/}, {\bf 130}(988), 065003.

\bibitem[{Jones} et~al.(2021){Jones}, {Martin}, {Anglada-Escud{\'e}},
  {Errmann}, {Campbell}, {Baker}, {Boonsri}, \&
  {Choochalerm}]{2021PASP..133b5001J}
{Jones}, H.~R.~A., {Martin}, W.~E., {Anglada-Escud{\'e}}, G., {Errmann}, R.,
  {Campbell}, D.~A., {Baker}, C., {Boonsri}, C., \& {Choochalerm}, P., 2021.
\newblock {A Small Actively Controlled High-resolution Spectrograph Based on
  Off-the-shelf Components}, {\it \pasp\/}, {\bf 133}(1020), 025001.

\bibitem[{Kawinkij} et~al.(2019){Kawinkij}, {Prasit}, {Buisset}, {Thummasorn},
  {Kuha}, {Lhospice}, {Jones}, {Martin}, {Errmann}, {Choochalerm},
  {Anglada-Escude}, {Campbell}, {Baker}, {Rattanasoon}, {Aukkaravittayapun},
  {Panyaphirawat}, {Leckngam}, {Mkrtichian}, {Poshyachinda}, \&
  {Soonthornthum}]{2019SPIE11116E..1GK}
{Kawinkij}, A., {Prasit}, A., {Buisset}, C., {Thummasorn}, G., {Kuha}, T.,
  {Lhospice}, E., {Jones}, H. R.~A., {Martin}, W.~E., {Errmann}, R.,
  {Choochalerm}, P., {Anglada-Escude}, G., {Campbell}, D., {Baker}, C.,
  {Rattanasoon}, S., {Aukkaravittayapun}, S., {Panyaphirawat}, T., {Leckngam},
  A., {Mkrtichian}, D., {Poshyachinda}, S., \& {Soonthornthum}, B., 2019.
\newblock {EXOhSPEC collimator mechanical design}, in {\em Astronomical Optics:
  Design, Manufacture, and Test of Space and Ground Systems II\/}, vol. 11116
  of {\bf Society of Photo-Optical Instrumentation Engineers (SPIE) Conference
  Series}, p. 111161G.

\bibitem[{Leon-Saval} et~al.(2005){Leon-Saval}, {Birks}, {Bland-Hawthorn}, \&
  {Englund}]{2005OptL...30.2545L}
{Leon-Saval}, S.~G., {Birks}, T.~A., {Bland-Hawthorn}, J., \& {Englund}, M.,
  2005.
\newblock {Multimode fiber devices with single-mode performance}, {\it Optics
  Letters\/}, {\bf 30}(19), 2545--2547.

\bibitem[{Lhospice} et~al.(2019){Lhospice}, {Buisset}, {Jones}, {Martin},
  {Errmann}, {Sithajan}, {Boonsri}, {Choochalerm}, {Anglada-Escud{\'e}},
  {Campbell}, {Alagao}, {Paenoi}, {Prasit}, {Panyaphirawat}, {Rattanasoon},
  {Leckngam}, {Mkrtichian}, \& {Aukkaravittayapun}]{2019SPIE11117E..0ZL}
{Lhospice}, E., {Buisset}, C., {Jones}, H. R.~A., {Martin}, W.~E., {Errmann},
  R., {Sithajan}, S., {Boonsri}, C., {Choochalerm}, P., {Anglada-Escud{\'e}},
  G., {Campbell}, D., {Alagao}, M.~A., {Paenoi}, J., {Prasit}, A.,
  {Panyaphirawat}, T., {Rattanasoon}, S., {Leckngam}, A., {Mkrtichian}, D., \&
  {Aukkaravittayapun}, S., 2019.
\newblock {EXOhSPEC folded design optimization and performance estimation}, in
  {\em Society of Photo-Optical Instrumentation Engineers (SPIE) Conference
  Series\/}, vol. 11117 of {\bf Society of Photo-Optical Instrumentation
  Engineers (SPIE) Conference Series}, p. 111170Z.

\bibitem[Liew et~al.(2016)Liew, Redding, Choma, Tagare, \& Cao]{Liew:16}
Liew, S.~F., Redding, B., Choma, M.~A., Tagare, H.~D., \& Cao, H., 2016.
\newblock Broadband multimode fiber spectrometer, {\it Opt. Lett.\/}, {\bf
  41}(9), 2029--2032.

\bibitem[{Mahadevan} et~al.(2012){Mahadevan}, {Ramsey}, {Bender}, {Terrien},
  {Wright}, {Halverson}, {Hearty}, {Nelson}, {Burton}, {Redman}, {Osterman},
  {Diddams}, {Kasting}, {Endl}, \& {Deshpande}]{2012SPIE.8446E..1SM}
{Mahadevan}, S., {Ramsey}, L., {Bender}, C., {Terrien}, R., {Wright}, J.~T.,
  {Halverson}, S., {Hearty}, F., {Nelson}, M., {Burton}, A., {Redman}, S.,
  {Osterman}, S., {Diddams}, S., {Kasting}, J., {Endl}, M., \& {Deshpande}, R.,
  2012.
\newblock {The habitable-zone planet finder: a stabilized fiber-fed NIR
  spectrograph for the Hobby-Eberly Telescope}, in {\em Ground-based and
  Airborne Instrumentation for Astronomy IV\/}, vol. 8446 of {\bf Society of
  Photo-Optical Instrumentation Engineers (SPIE) Conference Series}, p. 84461S.

\bibitem[{Mahadevan} et~al.(2014){Mahadevan}, {Halverson}, {Ramsey}, \&
  {Venditti}]{2014ApJ...786...18M}
{Mahadevan}, S., {Halverson}, S., {Ramsey}, L., \& {Venditti}, N., 2014.
\newblock {Suppression of Fiber Modal Noise Induced Radial Velocity Errors for
  Bright Emission-line Calibration Sources}, {\it \apj\/}, {\bf 786}(1), 18.

\bibitem[{Mayor} et~al.(2003){Mayor}, {Pepe}, {Queloz}, {Bouchy}, {Rupprecht},
  {Lo Curto}, {Avila}, {Benz}, {Bertaux}, {Bonfils}, {Dall}, {Dekker},
  {Delabre}, {Eckert}, {Fleury}, {Gilliotte}, {Gojak}, {Guzman}, {Kohler},
  {Lizon}, {Longinotti}, {Lovis}, {Megevand}, {Pasquini}, {Reyes}, {Sivan},
  {Sosnowska}, {Soto}, {Udry}, {van Kesteren}, {Weber}, \&
  {Weilenmann}]{2003Msngr.114...20M}
{Mayor}, M., {Pepe}, F., {Queloz}, D., {Bouchy}, F., {Rupprecht}, G., {Lo
  Curto}, G., {Avila}, G., {Benz}, W., {Bertaux}, J.~L., {Bonfils}, X., {Dall},
  T., {Dekker}, H., {Delabre}, B., {Eckert}, W., {Fleury}, M., {Gilliotte}, A.,
  {Gojak}, D., {Guzman}, J.~C., {Kohler}, D., {Lizon}, J.~L., {Longinotti}, A.,
  {Lovis}, C., {Megevand}, D., {Pasquini}, L., {Reyes}, J., {Sivan}, J.~P.,
  {Sosnowska}, D., {Soto}, R., {Udry}, S., {van Kesteren}, A., {Weber}, L., \&
  {Weilenmann}, U., 2003.
\newblock {Setting New Standards with HARPS}, {\it The Messenger\/}, {\bf 114},
  20--24.

\bibitem[{McCoy} et~al.(2012){McCoy}, {Ramsey}, {Mahadevan}, {Halverson}, \&
  {Redman}]{2012SPIE.8446E..8JM}
{McCoy}, K.~S., {Ramsey}, L., {Mahadevan}, S., {Halverson}, S., \& {Redman},
  S.~L., 2012.
\newblock {Optical fiber modal noise in the 0.8 to 1.5 micron region and
  implications for near infrared precision radial velocity measurements}, in
  {\em Ground-based and Airborne Instrumentation for Astronomy IV\/}, vol. 8446
  of {\bf Society of Photo-Optical Instrumentation Engineers (SPIE) Conference
  Series}, p. 84468J.

\bibitem[{M{\'e}gevand} et~al.(2012){M{\'e}gevand}, {Zerbi}, {Cabral}, {Di
  Marcantonio}, {Amate}, {Pepe}, {Cristiani}, {Rebolo}, {Santos}, {Dekker},
  {Abreu}, {Affolter}, {Avila}, {Baldini}, {Bristow}, {Broeg}, {Carvas},
  {Cirami}, {Coelho}, {Comari}, {Conconi}, {Coretti}, {Cupani}, {D'Odorico},
  {De Caprio}, {Delabre}, {Figueira}, {Fleury}, {Fragoso}, {Genolet}, {Gomes},
  {Gonzalez Hernandez}, {Hughes}, {Iwert}, {Kerber}, {Landoni}, {Lima},
  {Lizon}, {Lovis}, {Maire}, {Mannetta}, {Martins}, {Moitinho}, {Molaro},
  {Monteiro}, {Rasilla}, {Riva}, {Santana Tschudi}, {Santin}, {Sosnowska},
  {Sousa}, {Span{\`o}}, {Tenegi}, {Toso}, {Vanzella}, {Viel}, \& {Zapatero
  Osorio}]{2012SPIE.8446E..1RM}
{M{\'e}gevand}, D., {Zerbi}, F.~M., {Cabral}, A., {Di Marcantonio}, P.,
  {Amate}, M., {Pepe}, F., {Cristiani}, S., {Rebolo}, R., {Santos}, N.~C.,
  {Dekker}, H., {Abreu}, M., {Affolter}, M., {Avila}, G., {Baldini}, V.,
  {Bristow}, P., {Broeg}, C., {Carvas}, P., {Cirami}, R., {Coelho}, J.,
  {Comari}, M., {Conconi}, P., {Coretti}, I., {Cupani}, G., {D'Odorico}, V.,
  {De Caprio}, V., {Delabre}, B., {Figueira}, P., {Fleury}, M., {Fragoso}, A.,
  {Genolet}, L., {Gomes}, R., {Gonzalez Hernandez}, J., {Hughes}, I., {Iwert},
  O., {Kerber}, F., {Landoni}, M., {Lima}, J., {Lizon}, J.-L., {Lovis}, C.,
  {Maire}, C., {Mannetta}, M., {Martins}, C., {Moitinho}, A., {Molaro}, P.,
  {Monteiro}, M., {Rasilla}, J.~L., {Riva}, M., {Santana Tschudi}, S.,
  {Santin}, P., {Sosnowska}, D., {Sousa}, S., {Span{\`o}}, P., {Tenegi}, F.,
  {Toso}, G., {Vanzella}, E., {Viel}, M., \& {Zapatero Osorio}, M.~R., 2012.
\newblock {ESPRESSO: the ultimate rocky exoplanets hunter for the VLT}, in {\em
  Ground-based and Airborne Instrumentation for Astronomy IV\/}, vol. 8446 of
  {\bf Society of Photo-Optical Instrumentation Engineers (SPIE) Conference
  Series}, p. 84461R.

\bibitem[Mickelson \& Weierholt(1983)]{Mickelson:83}
Mickelson, A.~R. \& Weierholt, A., 1983.
\newblock Modal noise-limited signal-to-noise ratios in multimode optical
  fibers, {\it Appl. Opt.\/}, {\bf 22}(19), 3084--3089.

\bibitem[Minardi et~al.(2021)Minardi, Harris, \& Labadie]{Minardi2021}
Minardi, S., Harris, R.~J., \& Labadie, L., 2021.
\newblock Astrophotonics: astronomy and modern optics, {\it The Astronomy and
  Astrophysics Review\/}, {\bf 29}, 6.

\bibitem[Monerie \& Jeunhomme(1980)]{Monerie1980}
Monerie, M. \& Jeunhomme, L., 1980.
\newblock Polarization mode coupling in long single-mode fibres, {\it Optical
  and Quantum Electronics\/}, {\bf 12}, 449--461.

\bibitem[{Oliva} et~al.(2019){Oliva}, {Rainer}, {Tozzi}, {Sanna}, {Iuzzolino},
  \& {Brucalassi}]{2019A&A...632A..21O}
{Oliva}, E., {Rainer}, M., {Tozzi}, A., {Sanna}, N., {Iuzzolino}, M., \&
  {Brucalassi}, A., 2019.
\newblock {Experimental characterization of modal noise in multimode fibers for
  astronomical spectrometers}, {\it \aap\/}, {\bf 632}, A21.

\bibitem[{Petersburg} et~al.(2018){Petersburg}, {McCracken}, {Eggerman},
  {Jurgenson}, {Sawyer}, {Szymkowiak}, \& {Fischer}]{2018ApJ...853..181P}
{Petersburg}, R.~R., {McCracken}, T.~M., {Eggerman}, D., {Jurgenson}, C.~A.,
  {Sawyer}, D., {Szymkowiak}, A.~E., \& {Fischer}, D.~A., 2018.
\newblock {Modal Noise Mitigation through Fiber Agitation for Fiber-fed Radial
  Velocity Spectrographs}, {\it \apj\/}, {\bf 853}(2), 181.

\bibitem[{Pike} et~al.(2020){Pike}, {Beno{\^\i}t}, {MacLachlan}, {Harris},
  {Gris-S{\'a}nchez}, {Lee}, {Birks}, \& {Thomson}]{2020MNRAS.497.3713P}
{Pike}, F.~A., {Beno{\^\i}t}, A., {MacLachlan}, D.~G., {Harris}, R.~J.,
  {Gris-S{\'a}nchez}, I., {Lee}, D., {Birks}, T.~A., \& {Thomson}, R.~R., 2020.
\newblock {Modal noise mitigation for high-precision spectroscopy using a
  photonic reformatter}, {\it \mnras\/}, {\bf 497}(3), 3713--3725.

\bibitem[{Quirrenbach} et~al.(2012){Quirrenbach}, {Amado}, {Seifert},
  {S{\'a}nchez Carrasco}, {Mandel}, {Caballero}, {Mundt}, {Ribas}, {Reiners},
  {Abril}, {Aceituno}, {Alonso-Floriano}, {Ammler-von Eiff}, {Anglada-Escude},
  {Antona Jim{\'e}nez}, {Anwand-Heerwart}, {Barrado y Navascu{\'e}s},
  {Becerril}, {Bejar}, {Benitez}, {Cardenas}, {Claret}, {Colome},
  {Cort{\'e}s-Contreras}, {Czesla}, {del Burgo}, {Doellinger}, {Dorda},
  {Dreizler}, {Feiz}, {Fernandez}, {Galadi}, {Garrido}, {Gonz{\'a}lez
  Hern{\'a}ndez}, {Guardia}, {Guenther}, {de Guindos}, {Guti{\'e}rrez-Soto},
  {Hagen}, {Hatzes}, {Hauschildt}, {Helmling}, {Henning}, {Herrero}, {Huber},
  {Huber}, {Jeffers}, {Joergens}, {de Juan}, {Kehr}, {Klutsch}, {K{\"u}rster},
  {Lalitha}, {Laun}, {Lemke}, {Lenzen}, {Lizon}, {L{\'o}pez del Fresno},
  {L{\'o}pez-Morales}, {L{\'o}pez-Santiago}, {Mall}, {Martin},
  {Mart{\'\i}n-Ruiz}, {Mirabet}, {Montes}, {Morales}, {Morales Mu{\~n}oz},
  {Moya}, {Naranjo}, {Oreiro}, {P{\'e}rez Medialdea}, {Pluto}, {Rabaza},
  {Ramon}, {Rebolo}, {Reffert}, {Rhode}, {Rix}, {Rodler}, {Rodr{\'\i}guez},
  {Rodr{\'\i}guez L{\'o}pez}, {Rodr{\'\i}guez P{\'e}rez}, {Rodriguez Trinidad},
  {Rohloff}, {S{\'a}nchez-Blanco}, {Sanz-Forcada}, {Sch{\"a}fer}, {Schiller},
  {Schmidt}, {Schmitt}, {Solano}, {Stahl}, {Storz}, {St{\"u}rmer}, {Suarez},
  {Thiele}, {Ulbrich}, {Vidal-Dasilva}, {Wagner}, {Winkler}, {Xu}, {Zapatero
  Osorio}, \& {Zechmeister}]{2012SPIE.8446E..0RQ}
{Quirrenbach}, A., {Amado}, P.~J., {Seifert}, W., {S{\'a}nchez Carrasco},
  M.~A., {Mandel}, H., {Caballero}, J.~A., {Mundt}, R., {Ribas}, I., {Reiners},
  A., {Abril}, M., {Aceituno}, J., {Alonso-Floriano}, J., {Ammler-von Eiff},
  M., {Anglada-Escude}, G., {Antona Jim{\'e}nez}, R., {Anwand-Heerwart}, H.,
  {Barrado y Navascu{\'e}s}, D., {Becerril}, S., {Bejar}, V., {Benitez}, D.,
  {Cardenas}, C., {Claret}, A., {Colome}, J., {Cort{\'e}s-Contreras}, M.,
  {Czesla}, S., {del Burgo}, C., {Doellinger}, M., {Dorda}, R., {Dreizler}, S.,
  {Feiz}, C., {Fernandez}, M., {Galadi}, D., {Garrido}, R., {Gonz{\'a}lez
  Hern{\'a}ndez}, J., {Guardia}, J., {Guenther}, E., {de Guindos}, E.,
  {Guti{\'e}rrez-Soto}, J., {Hagen}, H.~J., {Hatzes}, A., {Hauschildt}, P.,
  {Helmling}, J., {Henning}, T., {Herrero}, E., {Huber}, A., {Huber}, K.,
  {Jeffers}, S., {Joergens}, V., {de Juan}, E., {Kehr}, M., {Klutsch}, A.,
  {K{\"u}rster}, M., {Lalitha}, S., {Laun}, W., {Lemke}, U., {Lenzen}, R.,
  {Lizon}, J.-L., {L{\'o}pez del Fresno}, M., {L{\'o}pez-Morales}, M.,
  {L{\'o}pez-Santiago}, J., {Mall}, U., {Martin}, E., {Mart{\'\i}n-Ruiz}, S.,
  {Mirabet}, E., {Montes}, D., {Morales}, J.~C., {Morales Mu{\~n}oz}, R.,
  {Moya}, A., {Naranjo}, V., {Oreiro}, R., {P{\'e}rez Medialdea}, D., {Pluto},
  M., {Rabaza}, O., {Ramon}, A., {Rebolo}, R., {Reffert}, S., {Rhode}, P.,
  {Rix}, H.-W., {Rodler}, F., {Rodr{\'\i}guez}, E., {Rodr{\'\i}guez L{\'o}pez},
  C., {Rodr{\'\i}guez P{\'e}rez}, E., {Rodriguez Trinidad}, A., {Rohloff},
  R.-R., {S{\'a}nchez-Blanco}, E., {Sanz-Forcada}, J., {Sch{\"a}fer}, S.,
  {Schiller}, J., {Schmidt}, C., {Schmitt}, J., {Solano}, E., {Stahl}, O.,
  {Storz}, C., {St{\"u}rmer}, J., {Suarez}, J.~C., {Thiele}, U., {Ulbrich}, R.,
  {Vidal-Dasilva}, M., {Wagner}, K., {Winkler}, J., {Xu}, W., {Zapatero
  Osorio}, M.~R., \& {Zechmeister}, M., 2012.
\newblock {CARMENES. I: instrument and survey overview}, in {\em Ground-based
  and Airborne Instrumentation for Astronomy IV\/}, vol. 8446 of {\bf Society
  of Photo-Optical Instrumentation Engineers (SPIE) Conference Series}, p.
  84460R.

\bibitem[{Ramsey}(1988)]{Ramsey:1988}
{Ramsey}, L.~W., 1988.
\newblock {Focal ratio degradation in optical fibers of astronomical
  interest.}, in {\em Fiber Optics in Astronomy\/}, vol.~3 of {\bf Astronomical
  Society of the Pacific Conference Series}, pp. 26--39.

\bibitem[{Raskin} et~al.(2020){Raskin}, {Rogozin}, {Mladenov}, {Schwab}, \&
  {Coutts}]{2020SPIE11203E..1OR}
{Raskin}, G., {Rogozin}, D., {Mladenov}, T., {Schwab}, C., \& {Coutts}, D.,
  2020.
\newblock {Fiber modal noise mitigation by a rotating double scrambler}, in
  {\em Advances in Optical Astronomical Instrumentation 2019\/}, vol. 11203 of
  {\bf Society of Photo-Optical Instrumentation Engineers (SPIE) Conference
  Series}, p. 112031O.

\bibitem[Redding et~al.(2014)Redding, Alam, Seifert, \& Cao]{Redding:14}
Redding, B., Alam, M., Seifert, M., \& Cao, H., 2014.
\newblock High-resolution and broadband all-fiber spectrometers, {\it
  Optica\/}, {\bf 1}(3), 175--180.

\bibitem[{Roy} et~al.(2014){Roy}, {Halverson}, {Mahadevan}, \&
  {Ramsey}]{2014SPIE.9147E..6BR}
{Roy}, A., {Halverson}, S., {Mahadevan}, S., \& {Ramsey}, L.~W., 2014.
\newblock {Scrambling and modal noise mitigation in the Habitable Zone Planet
  Finder fiber feed}, in {\em Ground-based and Airborne Instrumentation for
  Astronomy V\/}, vol. 9147 of {\bf Society of Photo-Optical Instrumentation
  Engineers (SPIE) Conference Series}, p. 91476B.

\bibitem[{Schwab} et~al.(2016){Schwab}, {Rakich}, {Gong}, {Mahadevan},
  {Halverson}, {Roy}, {Terrien}, {Robertson}, {Hearty}, {Levi}, {Monson},
  {Wright}, {McElwain}, {Bender}, {Blake}, {St{\"u}rmer}, {Gurevich},
  {Chakraborty}, \& {Ramsey}]{2016SPIE.9908E..7HS}
{Schwab}, C., {Rakich}, A., {Gong}, Q., {Mahadevan}, S., {Halverson}, S.~P.,
  {Roy}, A., {Terrien}, R.~C., {Robertson}, P.~M., {Hearty}, F.~R., {Levi},
  E.~I., {Monson}, A.~J., {Wright}, J.~T., {McElwain}, M.~W., {Bender}, C.~F.,
  {Blake}, C.~H., {St{\"u}rmer}, J., {Gurevich}, Y.~V., {Chakraborty}, A., \&
  {Ramsey}, L.~W., 2016.
\newblock {Design of NEID, an extreme precision Doppler spectrograph for WIYN},
  in {\em Ground-based and Airborne Instrumentation for Astronomy VI\/}, vol.
  9908 of {\bf Society of Photo-Optical Instrumentation Engineers (SPIE)
  Conference Series}, p. 99087H.

\bibitem[{Seifahrt} et~al.(2018){Seifahrt}, {St{\"u}rmer}, {Bean}, \&
  {Schwab}]{2018SPIE10702E..6DS}
{Seifahrt}, A., {St{\"u}rmer}, J., {Bean}, J.~L., \& {Schwab}, C., 2018.
\newblock {MAROON-X: a radial velocity spectrograph for the Gemini
  Observatory}, in {\em Ground-based and Airborne Instrumentation for Astronomy
  VII\/}, vol. 10702 of {\bf Society of Photo-Optical Instrumentation Engineers
  (SPIE) Conference Series}, p. 107026D.

\bibitem[{Spronck} et~al.(2012){Spronck}, {Fischer}, {Kaplan}, \&
  {Schwab}]{2012SPIE.8446E..8ZS}
{Spronck}, J.~F.~P., {Fischer}, D.~A., {Kaplan}, Z.~A., \& {Schwab}, C., 2012.
\newblock {Fiber scrambling for precise radial velocities at Lick and Keck
  Observatories}, in {\em Ground-based and Airborne Instrumentation for
  Astronomy IV\/}, vol. 8446 of {\bf Society of Photo-Optical Instrumentation
  Engineers (SPIE) Conference Series}, p. 84468Z.

\bibitem[{Spronck} et~al.(2013){Spronck}, {Fischer}, {Kaplan}, {Schwab}, \&
  {Szymkowiak}]{2013PASP..125..511S}
{Spronck}, J. F.~P., {Fischer}, D.~A., {Kaplan}, Z.~A., {Schwab}, C., \&
  {Szymkowiak}, A., 2013.
\newblock {Fiber Scrambling for High-Resolution Spectrographs. I. Lick
  Observatory}, {\it \pasp\/}, {\bf 125}(927), 511.

\bibitem[{Spronck} et~al.(2015){Spronck}, {Fischer}, {Kaplan}, {Jurgenson},
  {Valenti}, {Moriarty}, \& {Szymkowiak}]{2015PASP..127.1027S}
{Spronck}, J. F.~P., {Fischer}, D.~A., {Kaplan}, Z., {Jurgenson}, C.~A.,
  {Valenti}, J., {Moriarty}, J., \& {Szymkowiak}, A.~E., 2015.
\newblock {Fiber Scrambling for High-Resolution Spectrographs. II. A Double
  Fiber Scrambler for Keck Observatory}, {\it \pasp\/}, {\bf 127}(956), 1027.

\bibitem[{Stoehr} et~al.(2008){Stoehr}, {White}, {Smith}, {Kamp}, {Thompson},
  {Durand}, {Freudling}, {Fraquelli}, {Haase}, {Hook}, {Kimball}, {K{\"u}mmel},
  {Levay}, {Lombardi}, {Micol}, \& {Rogers}]{2008ASPC..394..505S}
{Stoehr}, F., {White}, R., {Smith}, M., {Kamp}, I., {Thompson}, R., {Durand},
  D., {Freudling}, W., {Fraquelli}, D., {Haase}, J., {Hook}, R., {Kimball}, T.,
  {K{\"u}mmel}, M., {Levay}, K., {Lombardi}, M., {Micol}, A., \& {Rogers}, T.,
  2008.
\newblock {DER\_SNR: A Simple \& General Spectroscopic Signal-to-Noise
  Measurement Algorithm}, in {\em Astronomical Data Analysis Software and
  Systems XVII\/}, vol. 394 of {\bf Astronomical Society of the Pacific
  Conference Series}, p. 505.

\bibitem[Yerolatsitis et~al.(2017)Yerolatsitis, Harrington, \&
  Birks]{Yerolatsitis:17}
Yerolatsitis, S., Harrington, K., \& Birks, T.~A., 2017.
\newblock All-fibre pseudo-slit reformatters, {\it Opt. Express\/}, {\bf
  25}(16), 18713--18721.

\end{thebibliography}








\bsp	
\label{lastpage}
\end{document}